\newif\iffigs\figstrue
\documentclass[a4paper,11pt]{article}
\usepackage{latexsym,amssymb,lscape,graphics}
\usepackage{graphicx}        % pacchetto grafico standard LaTeX
\usepackage{longtable}
\usepackage{multirow}
\usepackage{color}
\usepackage{slashed,epsfig}
\usepackage{amsfonts}
\newtheorem{definizione}{Definition}[section]

\newcommand{\bd}{\begin{definizione}}
\newcommand{\ed}{\end{definizione}}
\def\IC{\relax\,\hbox{$\inbar\kern-.3em{\rm C}$}}
\def\IG{\relax\,\hbox{$\inbar\kern-.3em{\rm G}$}}
\def\IB{\relax{\rm I\kern-.18em B}}
\def\ID{\relax{\rm I\kern-.18em D}}
\def\IL{\relax{\rm I\kern-.18em L}}
\def\IF{\relax{\rm I\kern-.18em F}}
\def\IH{\relax{\rm I\kern-.18em H}}
\def\II{\relax{\rm I\kern-.17em I}}
\def\IN{\relax{\rm I\kern-.18em N}}
\def\IP{\relax{\rm I\kern-.18em P}}
\def\IQ{\relax\,\hbox{$\inbar\kern-.3em{\rm Q}$}}
\def\bfzero{\relax\,\hbox{$\inbar\kern-.3em{\rm 0}$}}
\def\IK{\relax{\rm I\kern-.18em K}}
\def\IG{\relax\,\hbox{$\inbar\kern-.3em{\rm G}$}}
 \font\cmss=cmss10 \font\cmsss=cmss10 at 7pt
\def\IR{\relax{\rm I\kern-.18em R}}
\def\ZZ{\relax\ifmmode\mathchoice
{\hbox{\cmss Z\kern-.4em Z}}{\hbox{\cmss Z\kern-.4em Z}}
{\lower.9pt\hbox{\cmsss Z\kern-.4em Z}} {\lower1.2pt\hbox{\cmsss
Z\kern-.4em Z}}\else{\cmss Z\kern-.4em Z}\fi}
\def\bfone{\relax{\rm 1\kern-.35em 1}}

\def\inbar{\vrule height1.5ex width.4pt depth0pt}
\def\bfzero{\relax{\rm I\kern-.18em 0}}
\def\bfone{\relax{\rm 1\kern-.35em 1}}

\DeclareFontFamily{U}{rsf}{} \DeclareFontShape{U}{rsf}{m}{n}{
  <5> <6> rsfs5 <7> <8> <9> rsfs7 <10-> rsfs10}{}
\DeclareMathAlphabet\Scr{U}{rsf}{m}{n}

\newcommand{\ft}[2]{{\textstyle\frac{#1}{#2}}}
\def\tilde{\widetilde}

\def\1bar{1\hskip -.275cm -}
\def\2bar{2\hskip -.275cm -}
\def\3bar{3\hskip -.275cm -}

\newsavebox{\uuunit}
\sbox{\uuunit}
                 {\setlength{\unitlength}{0.825em}
                      \begin{picture}(0.6,0.7)
                                      \thinlines
                                      \put(0,0){\line(1,0){0.5}}
                                      \put(0.15,0){\line(0,1){0.7}}
                                      \put(0.35,0){\line(0,1){0.8}}
                                     \multiput(0.3,0.8)(-0.04,-0.02){10}{\rule{0.5pt}{0.5pt}}
                      \end {picture}}

\makeatletter \@addtoreset{equation}{section} \makeatother
\def\bfone{\relax{\rm 1\kern-.35em 1}}

\def\bfone{\relax{\rm 1\kern-.35em 1}}
\font\cmss=cmss10 \font\cmsss=cmss10 at 7pt
\newcommand{\uu}{\mathfrak{u}}

\def\IE{\relax{{\rm I\kern-.18em E}}}

\def\IGam{\relax{{\rm I}\kern-.18em \Gamma}}

\def\IA{\relax{\hbox{{\rm A}\kern-.82em {\rm A}}}}

\begin{document}
\begin{titlepage}

{\vspace{-2.8em}}

\hfill {CERN-PH-TH/2013-274}
%\sk

\vskip 0.4cm
\begin{center}
{\Large {\sc  On the Topology of the  Inflaton Field \\
in \\
Minimal Supergravity Models}}\\[1cm]
{ \large Sergio Ferrara$^{a}$, Pietro Fr\'e$^{b}$\footnote{Prof. Fr\'e is presently fulfilling the duties of Scientific Counselor of the Italian Embassy in the Russian Federation, Denezhnij pereulok, 5, 121002 Moscow, Russia.}, Alexander S. Sorin$^{c}$ }
{}~\\
{}~\\
\quad
\quad \\{{\em $^a$ Physics Department, Theory Unit, CERN, CH 1211, Geneva 23, Switzerland, {\tt and}\\
INFN - Laboratori Nazionali di Frascati, Via Enrico Fermi 40, I-00044, Frascati, Italy}}, {\tt and}\quad \\
Department of Physics and Astronomy, University of California, Los Angeles, CA 90095-1547, USA~\quad \\
{\tt sergio.ferrara@cern.ch}
{}~\\
\quad \\
{{\em $^{b}$  Dipartimento di Fisica, Universit\'a di Torino,}}
\\
{{\em $\&$ INFN - Sezione di Torino}}\\
{\em via P. Giuria 1, I-10125 Torino, Italy}~\quad\\
{\tt fre@to.infn.it}
{}~\\
\quad \\
{{\em $^{c}$ Bogoliubov Laboratory of Theoretical Physics, {\tt and} }}\\
{{\em  Veksler and Baldin Laboratory of High Energy Physics,}}\\
{{\em Joint Institute for Nuclear Research,}}\\
{\em 141980 Dubna, Moscow Region, Russia}~\quad\\
{\tt sorin@theor.jinr.ru}
\quad \\
\end{center}
~{}
\begin{abstract}
We consider global issues in minimal supergravity models where a single
field inflaton potential emerges. In a particular case we reproduce the Starobinsky model and its description dual to a certain formulation
of $R+R^2$ supergravity. For definiteness we confine our analysis to spaces at constant curvature, either vanishing or negative. Five distinct models arise, two flat models with respectively a quadratic and a quartic potential and three based on the $\frac{\mathrm{SU(1,1)}}{\mathrm{U(1)}}$ space where its distinct isometries, elliptic, hyperbolic and parabolic are gauged. Fayet-Iliopoulos terms are introduced in a geometric way and they turn out to be a crucial ingredient in order to describe the de Sitter inflationary phase of the Starobinsky model.
\end{abstract}
\end{titlepage}
\tableofcontents
\newpage
\section{Introduction}
The accurate results on the CMB power spectrum collected firstly by the WMAP mission and more recently by the PLANCK satellite \cite{Ade:2013uln}\cite{Ade:2013zuv},\cite{Hinshaw:2012aka} have boosted a new wave of research activities on the theoretical modelling of the inflationary paradigma and seem to favour the scenario based on a single scalar field $\phi$ (the inflaton) with a suitable potential $V(\phi)$. In the notations adopted in the present paper, the Friedman equations that govern the time evolution of the scale factor $a(t)$ and of the inflaton $\phi(t)$ are written as follows:
\begin{equation}% \nonumber to remove numbering (before each equation)
 H^2 \ = \  \frac{1}{3} \ \dot{\phi}^2 \, + \, \frac{2}{3} \ V(\phi)  \quad ; \quad  \dot{H}  \ = \ - \, \dot{\phi}^2   \quad ; \quad \ddot{\phi} \,+ \,  3 \, H \, \dot{\phi} \, + \, V^{\,\prime}  \ = \ 0  \label{fridmano}
\end{equation}
where
$ H(t) \, \equiv \, \frac{\dot{a}(t)}{a(t)}$ is the Hubble function. Equations (2.1) and (2.2) of the recent review \cite{Encyclopaedia} of inflationary models coincide with eq.s (\ref{fridmano})  if one chooses the convention $2 \, M_{Pl}^2 \, = \, 1$. This observation, together with the statement that the kinetic term of the inflaton is canonical in our lagrangian :
\begin{equation}\label{gargamello}
  \mathcal{L}\, = \, \dots + \, \ft 12 \, \partial_\mu \phi \partial^\mu \phi \, + \, \dots
\end{equation}
 fixes completely all normalizations and allows the comparison of the results we shall present here with any other result  in  the vast literature on inflation.
\par
After the publication of PLANCK data, the issue whether one inflaton cosmological models with realistic potentials could be embedded into $\mathcal{N}=1$ supergravity in a minimal way was addressed and resolved in a series of recent papers
\cite{minimalsergioKLP},\cite{Ferrara:2013wka},\cite{Ferrara:2013kca}.  Any inflation model based on a positive definitive potential can be embedded into $\mathcal{N}=1$ supergravity, coupled to a single Wess-Zumino multiplet and one massless vector multiplet, which may combine together in a massive vector multiplet with lagrangian specified by a single real function J(C) as shown in
\cite{VanProeyen:1979ks}.
%zdes'
The vector multiplet is utilized to gauge an isometry of the one-dimensional Hodge-K\"ahler manifold $\Sigma$ associated with the WZ multiplet. The catch of the method is the supergravity formulation of the Higgs mechanism. The gauging introduces a $D$-term and the definite potential $V$ is the square of the momentum-map of the Killing vector $k^\mathfrak{z}$ that generates the gauged isometry of $\Sigma$. One of the two scalar fields of the WZ multiplet, conventionally named $B$, is eaten up by the vector field $\mathcal{A}_\mu$ which becomes massive and the other one $C$, after a field transformation $C\to \phi(C)$  that reduces it to have the canonical form (\ref{gargamello}), becomes the inflaton.
We name $C(\phi)$ the Van Proeyen coordinate on the K\"ahler manifold $\Sigma_K$ since it corresponds to the scalar field in terms of which the  $\mathcal{N}=1$ supergravity lagrangian that turns out to be the minimal one for inflationary models was firstly written by Van Proeyen in \cite{VanProeyen:1979ks}. In the sequel we shall emphasize the intrinsic geometrical meaning of the VP coordinate $C$. The minimal models for
inflation of \cite{minimalsergioKLP} were suggested  by the supergravity completion \cite{Cecotti:1987qe} of
the $R+R^2$ Starobinski model which, as we will further discuss hereby, corresponds
manifold $\Sigma$ of constant curvature $\nu^2 \, = \, \ft 43$ (see eq.(\ref{formidabile})) with gauged shift symmetry and and a non vanishing Fayet Iliopoulos term \footnote{The reader should notice that the field $\phi$ used in this paper differs from the field $\tilde{\phi}$ used in many other papers by a factor $\frac{1}{\sqrt{2}}$ as it is evident from the normalization of Friedman equations in eq.(\ref{fridmano}). Correspondingly the exponential $\exp[\tilde{{\nu}} \tilde{\phi}]$ in other paper normalizations is $\exp[ \sqrt{2} \tilde{\nu} \phi]$ in our present normalizations. In this way the Starobinsky potential corresponds to $\nu^2 \, = \, 4/3$ in our normalizations if it corresponds to $\nu^2 \, = \, 2/3$ in other paper normalizations.}
\par
In the same period, following work on the phenomenon of climbing scalars \cite{Dudasprimo},\cite{Dudas:2012vv},\cite{Sagnotti:2013ica}, another series of papers  \cite{noicosmoitegr,mariosashapietrocosmo}  addressed the issue of integrability of the two field\footnote{The two fields are the scale factor $a(t)$ and the inflaton $\phi(t)$.} dynamical system encoded in the Friedman equations (\ref{fridmano}) and provided a list of 28 integrable potentials $V(\phi)$ in the sense that they provide  integrabity of eq.s (\ref{fridmano}). The question if any of these integrable $V(\phi)$ can be embedded into gauged extended supergravity was discussed in \cite{mariosashapietrocosmo} and remains partially open, yet as a consequence of the results of  \cite{minimalsergioKLP}, the embedding of all positive definite ones among them  into $\mathcal{N}=1$ minimal supergravity is guaranteed and deserves careful considerations for the possibility that this discloses, within a supergravity context of basing  Mukhanov-Sasaki equations on exact analytic solutions of the Klein Gordon Einstein system.
\par
Independently from integrability, the geometrical basis of the construction of minimal supergravity models  of inflation introduced in \cite{minimalsergioKLP} was analyzed in another pair of recent papers \cite{primosashapietro},\cite{piesashatwo}. It was pointed out that the root from any given positive definite potential $V(\phi)$ to the corresponding minimal supergravity model is a map, named by two of us the $D$-map, in whose image there is a two-dimensional K\"ahler surface $\Sigma$ admitting at least a one-dimensional group of isometries $\mathcal{G}$. Various aspects of this map were explored, but a fundamental question remained so far unanswered about the global topology both of the surface $\Sigma$ and of its isometry group. This is by no means a marginal issue. Indeed, as we are going to show here, the physical properties and the symmetries of the minimal supergravity lagrangian are significantly different in the two cases of a compact isometry group $\mathcal{G}\, = \, \mathrm{U(1)}$ and of a non compact one $\mathcal{G}\, = \, \mathrm{SO(1,1)}$ or $\mathcal{G}\, = \, \mathbb{R}$. Furthermore the asymptotic behavior of the real function $J(C)$ that defines both the potential and the kinetic terms of the scalars,  is distinct in the case of compact and non compact symmetries and actually provides a clue to identify the appropriate global topology. In this paper we exemplify these concepts by classifying all minimal supergravity models and corresponding inflaton potentials that are associated with a K\"ahler surface $\Sigma$ of constant curvature $R_\Sigma$. We obtain a total of five models, each still depending on one or two parameters, that are associated with $R_\Sigma \, = \, 0$ (flat models) and with $R_\Sigma \, = \, - \, 4 \, \nu^2$. In the latter case the corresponding manifold is always $\Sigma \, = \, \mathrm{SL(2,\mathbb{R})/O(2)}$, but by gauging elliptic, hyperbolic or parabolic subgroups we obtain different  families of potentials. The Starobinsky type of potentials \cite{Starobinsky:1980te} are obtained from the parabolic subgroups. Various discussions of Starobinsky potentials  and other inflaton potentials within supergravity have been advanced in other papers \cite{johndimitri},\cite{Ketov:2010qz},\cite{Ketov:2012jt},\cite{Kallosh:2013maa},\cite{Kallosh:2013daa},
\cite{Kallosh:2013tua},\cite{Kallosh:2013yoa},\cite{Kallosh:2013hoa},\cite{Kallosh:2013lkr},\cite{Farakos:2013cqa}.
\section{Global structure of the inflaton K\"ahler surface}
\label{axialsurf}
As we advocated in the introduction, in the minimal $\mathcal{N}=1$ supergravity realizations of one--scalar cosmologies the central item of the construction is an axial(-shift) symmetric K\"ahler surface whose metric can be written as follows:
\begin{equation}\label{metraxia}
  ds^2_\Sigma \, = \, p(U) \, dU^2 \, + \, q(U) \, dB^2
\end{equation}
$p(U),q(U)$ being two positive definite functions of their argument. The manifold $\Sigma$ is an axial(-shift) symmetric surface, since the metric (\ref{metraxia}) admits the Killing vector $\vec{k}_{[B]} \, = \, \partial_B$. This isometry is fundamental since it is by means of its gauging that one produces a $D$-type positive definite scalar potential that can encode the inflaton dynamics.  At the level of the supergravity model that is built by using the K\"ahler space $\Sigma$ as the target manifold where  the two scalar fields of the inflatonic Wess Zumino multiplet take values, a fundamental question is whether $\vec{k}_{[B]}$ generates a \textbf{compact rotation symmetry} or a \textbf{non compact shift symmetry}. Indeed the supergravity lagrangian in general and its fermionic sector in particular,  display quite different features in the two cases, leading to a different pattern of physical charges and symmetries. Actually, as we are going to illustrate below, when $\Sigma \, = \, \Sigma_{max}$ is a constant curvature surface namely the coset manifold $\frac{\mathrm{SU(1,1)}}{\mathrm{U(1)}}\sim \frac{\mathrm{SL(2,\mathbb{R})}}{\mathrm{O(2)}}$, there is also a third possibility. In such a situation the killing vector  $\vec{k}_{[B]}$ can be the generator of a \textbf{dilatation}, namely it can correspond to a non-compact but semi-simple element $\mathbf{d}\, = \, \left(
                                                                                 \begin{array}{cc}
                                                                                   1 & 0 \\
                                                                                   0 & -1 \\
                                                                                 \end{array}
                                                                               \right)
$ of the Lie algebra $\mathrm{SL(2,\mathbb{R})}$ rather then to a nilpotent one $\mathbf{t} \, = \,\left(
                                     \begin{array}{cc}
                                       0 & 1 \\
                                       0 & 0 \\
                                     \end{array}
                                   \right)
$.
In the ambient algebra $\mathfrak{g}\, = \,\mathrm{SL(2,\mathbb{R})}$ this distinction makes sense since by means of internal transformations (conjugations  $\mathbf{d}^\prime \, = \, \exp[-\mathbf{u}] \, \mathbf{d} \, \exp[\mathbf{u}]$ with $\mathbf{u} \in \mathfrak{g}$) we cannot map $\mathbf{d}$ into $\mathbf{t}$.  The main question therefore  concerns  the global topology of such a group. Is it compact
$\mathcal{G} \sim \mathrm{U(1)}$ or is it non-compact $\mathcal{G} \sim \mathbb{R}$? As already advocated, in the two cases the structure of the $\mathcal{N}=1$ supergravity lagrangian is different and its local and global symmetries are different.
\par
As it was explained in \cite{piesashatwo}, the standard presentation of the geometry of $\Sigma$ in terms of a complex coordinate and of a K\"ahler potential is obtained by means of a few standard steps. First one singles out the unique complex structure with vanishing Nienhuis tensor with respect to which the metric is hermitian:
\begin{equation}\label{crisanto}
   \mathfrak{J}_\alpha^\beta \,\mathfrak{ J}_\beta^\gamma \, = \, - \, \delta^\gamma_\alpha \quad ;\quad \partial_{[\alpha } \, \mathfrak{J}^\gamma_{\beta]} \, - \, \mathfrak{J}^\mu_\alpha \,
  \mathfrak{J}^\nu_\beta \,  \partial_{[\mu } \, \mathfrak{J}^\gamma_{\nu]} \, = \, 0 \quad ; \quad  g_{\alpha\beta} \, = \, \mathfrak{ J}_\alpha^\gamma \, \mathfrak{ J}_\beta^\delta \, g_{\gamma\delta}
\end{equation}
In terms of the metric coefficients, such a complex structure is  given by the following tensor $\mathfrak{J}$ and leads to the following closed K\"ahler 2-form $\mathrm{K}$:
\begin{equation}\label{forbito}
 \mathfrak{ J} \, = \, \left( \begin{array}{cc} 0 & \sqrt{\frac{p(U)}{q(U)}} \\
                              - \, \sqrt{\frac{q(U)}{p(U)}} & 0 \\
                     \end{array} \right) \quad \Rightarrow \quad \mathrm{K} \, = \, g_{\alpha\mu}\,\mathfrak{J}^\mu_\beta \, \, dx^\alpha \, \wedge \, dx^\beta \, =\, - \, \sqrt{p(U) \, q(U) } \, dU \, \wedge \, dB
\end{equation}
Next one aims at reproducing the K\"ahlerian metric (\ref{metraxia}) in terms of a complex coordinate $\mathfrak{z}\, = \,\mathfrak{z}(U,B) $  and a K\"ahler potential $\mathcal{K}(\mathfrak{z} \, , \, \bar{\mathfrak{z}})\, = \,\mathcal{K}^\star(\mathfrak{z} \, , \, \bar{\mathfrak{z}}) $ such that:
\begin{equation}\label{foxterry}
  \mathrm{K} \, = \, \partial \, \overline{\partial} \, \mathcal{K} \, = \, {\rm i} \partial_{\mathfrak{z}} \, \partial_{\bar{\mathfrak{z}}} \, \mathcal{K} \, d\mathfrak{z} \, \wedge \, d\bar{\mathfrak{z}} \quad ; \quad ds^2_{\Sigma} \, = \, \partial_{\mathfrak{z}} \, \partial_{\bar{\mathfrak{z}}} \, \mathcal{K} \, d\mathfrak{z} \, \otimes \, d\bar{\mathfrak{z}}
\end{equation}
As explained in \cite{piesashatwo} the complex coordinate $\mathfrak{z}$ is necessarily a solution of the complex structure equation:
\begin{equation}\label{golosina}
  \mathfrak{ J}_\alpha^\beta \, \partial_\beta \, \mathfrak{z} \, = \, {\rm i} \partial_\alpha \, \mathfrak{z} \quad \Rightarrow \quad \sqrt{\frac{p(U)}{q(U)}} \, \partial_B \, \mathfrak{z}(U,B) \, = \, {\rm i} \, \partial_U \, \mathfrak{z}(U,B)
\end{equation}
The general solution of such an equation is easily found. Define the linear combination \footnote{As it follows from the present discussion the Van Proeyen coordinate $C(U)$ has an intrinsic geometric characterization as that one which solves the differential equation of the complex structure.} :
\begin{equation}\label{gomorra}
    w \, \equiv \, {\rm i} \, C(U) \, - \, B \quad ; \quad C(U) \, = \, \int \, \sqrt{\frac{p(U)}{q(U)}} \, dU
\end{equation}
and consider any holomorphic function $f(w)$. As one can immediately verify, the position:
\begin{equation}\label{frantastic}
    \mathfrak{z}(U,B) \, = \, f(w)
\end{equation}
solves eq.(\ref{golosina}). What is the appropriate choice of the holomorphic function $f(w)$? Locally (in an open neighborhood) this is an empty question, since  the holomorphic function $f(t)$ simply corresponds to a change of coordinates and gives rise to the same K\"ahler metric in a different basis. Suppose we have selected a particular function $f(w)$ and setting $\mathfrak{z} \, = \, f(w)$ we have found a K\"ahler function $\mathcal{K}(\mathfrak{z},\bar{\mathfrak{z}})$ such that:
\begin{equation}\label{gorilla}
    ds^2 \, \equiv \, \partial_\mathfrak{z} \, \partial_{\bar{\mathfrak{z}}} \, \mathcal{K} \, d\mathfrak{z} \, d\bar{\mathfrak{z}} \, = \, p(U) \, dU^2 \, + \, q(U) \, dB^2
\end{equation}
then, by performing the holomorphic transformation $\hat{\mathfrak{z}}\, = \, g(\mathfrak{z}) \, = \, g\left(f(w)\right) \, = \, \hat{f}(w)$ we obtain a locally equivalent presentation of the same metric. Writing
$\hat{\mathcal{K}}(\hat{\mathfrak{z}},\hat{\bar{\mathfrak{z}}}) \, = \, \mathcal{K}\left(g^{-1}(\hat{\mathfrak{z}}),{\bar{g}}^{-1}(\hat{\bar{\mathfrak{z}}})\right)$ we obviously get:
\begin{equation}\label{gorillaTwo}
    ds^2 \, \equiv \, \partial_{\hat{\mathfrak{z}}} \, \partial_{\hat{\bar{\mathfrak{z}}}} \, \hat{\mathcal{K}} \, d\hat{\mathfrak{z}} \, d\hat{\bar{\mathfrak{z}}} \, = \, p(U) \, dU^2 \, + \, q(U) \, dB^2
\end{equation}
Globally, however, there are significant restrictions that concern the range of the variables $B$ and $C(U)$, namely the global topology of the manifold $\Sigma$. By definition $B$ is the coordinate that, within $\Sigma$, parameterizes  points along the $\mathcal{G}$-orbits. If $\mathcal{G}$ is compact, then $B$ is a  coordinate on the circle and it must be defined up to identifications $B\simeq B + 2\, n \, \pi $, where $n$ is an integer. On the other hand if $B$ is non compact its range extends on the full real line $\mathbb{R}$. Furthermore, in order to obtain a presentation of the K\"ahler geometry of $\Sigma$ that allows to single out a \textit{canonical inflaton} field $\phi$ with a potential $V(\phi)$  we aim at a K\"ahler potential $\mathcal{K}(\mathfrak{z},\bar{\mathfrak{z}})$ that in terms of the variables $C(U)$ and $B$ should actually depend only on $C$ being constant on the $\mathcal{G}$-orbits. Starting from the metric (\ref{metraxia}) we can always choose a canonical variable $\phi$ defined by the position:
 \begin{equation}\label{gonzallo}
  \phi \, = \, \phi(U) \, = \, \int \, 2 \,  \sqrt{p(U)} \, dU \quad ; \quad \ft 12 \, d\phi \, = \, \sqrt{p(U)} \, dU
\end{equation}
and assuming that $\phi(U)$ can be inverted $U \, = \, U(\phi)$ we can rewrite (\ref{metraxia}) in the following canonical form:
\begin{equation}\label{solarium}
  ds^2_{can} \, = \, \ft 14 \, \left[ d\phi^2 \, + \, \left(\mathcal{ P}^\prime(\phi)\right)^2 \, dB^2 \right] \quad ; \quad
  \mathcal{ P}^\prime(\phi) \, = \, 2 \, \sqrt{ q\left(U(\phi)\right)} \quad ; \quad \underbrace{\sqrt{p(U(\phi))} \, \frac{dU}{d\phi} \, = \, \ft 12}_{\mbox{by construction}}
  \end{equation}
The reason to call the square root of $q\left(U(\phi)\right)$ with the name $\mathcal{ P}^\prime(\phi)$  is the  interpretation  of such a function as the derivative with respect to the canonical variable $\phi$ of the momentum map of the Killing vector  $\vec{k}_{[B]}$. As it was pointed out in \cite{piesashatwo} such interpretation is crucial  for the construction of the corresponding supergravity model but it is intrinsic to the geometry of the surface $\Sigma$.
\par
According to an analysis first introduced in section 4 of \cite{minimalsergioKLP}, by using the canonical variable $\phi$, the VP coordinate $C$ defined in equation (\ref{gomorra}) becomes:
\begin{equation}\label{sodoma}
    C(\phi) \, = \, C\left(U(\phi)\right) \, = \, \int \, \frac{d\phi}{\mathcal{ P}^\prime(\phi)}
\end{equation}
and the metric $ds^2_{\Sigma} \, = \, ds^2_{can}$ of the K\"ahler surface $\Sigma$  can be rewritten as:
\begin{equation}\label{Jmet}
   ds^2_{\Sigma} \, = \, \ft 14 \, \frac{\mathrm{d}^2J}{\mathrm{d}C^2} \left( \mathrm{d}C^2 \, + \mathrm{d}B^2\right)
\end{equation}
where the function $J(C)$ is defined as follows\footnote{See \cite{piesashatwo} for more details.}:
\begin{equation}
    \mathcal{J}(\phi) \, \equiv \, \int \, \frac{\mathcal{P}(\phi)}{\mathcal{P}^\prime(\phi)} \, d\phi \quad ; \quad J(C) \, \equiv \,  \mathcal{J}\left(\phi(C)\right) \label{gartoccio}
\end{equation}
It appears from the above formula that the crucial step in working out the analytic form of the function $J(C)$ is the ability of inverting the relation between the VP coordinate $C$, defined by the integral (\ref{sodoma}), and the canonical one $\phi$, a task which, in the general case, is quite hard in both directions. The indefinite integral (\ref{sodoma}) can be expressed in terms of special functions only in certain cases and even less frequently one has at his own disposal  inverse functions. Yet this is only a technical difficulty. Conceptually, eq.s(\ref{gartoccio}) and (\ref{sodoma}) define the function $J(C)$ up to an additive integration constant. The fundamental unanswered question is how to reinterpret eq.(\ref{Jmet}) in terms of a complex coordinate $\mathfrak{z}$ and of a K\"ahler potential $\mathcal{K}(\mathfrak{z},\bar{\mathfrak{z}})$. Having already established in eq.(\ref{gomorra}) the general solution of the complex structure equations there are three possibilities that correspond, in the case of constant curvature manifolds $\Sigma_{max}$  to the three conjugacy classes of $\mathrm{SL(2,\mathbb{R})}$ elements (elliptic, hyperbolic and parabolic). In the three cases $J(C)$ is identified with the K\"ahler potential $\mathcal{K}(\mathfrak{z},\bar{\mathfrak{z}})$, but it remains to be decided whether the VP coordinate $C$ is to be identified with the imaginary part of the complex coordinate $C \, = \, \mbox{Im} \, \mathfrak{z}$,  with the logarithm of its modulus $C \, = \, \ft 12 \, \log \, |\mathfrak{z}|^2$, or with a third combination of $\mathfrak{z}$ and $\bar{\mathfrak{z}}$, namely whether we choose the first  the second or the third  of the options listed below:
\begin{equation}\label{curlandia}
 \mathfrak{z} \, = \, \left \{ \begin{array}{rccclc}
                                 \zeta & \equiv & \exp\left[\,-\,{\rm i}\, w\right] &=&\underbrace{\exp\left[C(\phi)\right]}_{\rho(\phi)} \, \exp\left[ {\rm i} B\right] \\
                                 t & \equiv &  w &=& {\rm i} \, C(\phi) - B \\
                                 \hat{\zeta} & \equiv & {\rm i} \, \tanh \left(- \,\ft 12 \, w \right)&=& {\rm i} \, \tanh \left(-\,\ft 12  \, ({\rm i} \, C(\phi) - B )\right)\\
                               \end{array} \right| \quad \quad C(\phi) \, \equiv \, \int \,  \frac{1}{\mathcal{P}^\prime(\phi)}  \, d\phi
\end{equation}
If we choose the first solution $\mathfrak{z} \, = \, \zeta$, that in \cite{piesashatwo} was named  of \textbf{Disk-type}, we obtain that the basic isometry generated by the Killing vector $\vec{k}_{[B]}$ is a compact rotation symmetry and this implies a series of consequences on the supergravity lagrangian and its symmetries that we discuss below. Choosing  the second solution $\mathfrak{z} \, = \, t$, that was named  of \textbf{Plane-type} in \cite{piesashatwo}, is appropriate instead to the case of a non compact shift symmetry and leads to different symmetries of the supergravity lagrangian. The third possibility mentioned above which occurs in the case of constant curvature surfaces $\Sigma_{max}$ and leads to the interpretation of the $B$-shift as an $\mathrm{SO(1,1)}$-hyperbolic transformation.
\par
In the three cases the analytic form of the holomorphic Killing vector  $\vec{k}_{[B]}$ is quite different:
\begin{equation}\label{familione}
    \vec{k}_{[B]} \, = \, \left \{ \begin{array}{lclccclcl}
                                     {\rm i} \zeta \, \partial_\zeta & \equiv &k^{\mathfrak{z}}\partial_{\mathfrak{z}} & \Rightarrow & k^{\mathfrak{z}} & = & {\rm i} \, \mathfrak{z} & ; & \mbox{Disk-type, compact rotation} \\
                                     \partial_t& \equiv &k^{\mathfrak{z}}\partial_{\mathfrak{z}}& \Rightarrow & k^{\mathfrak{z}} & = & 1 & ; & \mbox{Plane-type, non-compact shift}\\
                                     {\rm i} \left(1+\hat{\zeta}^2\right) \partial_{\hat{\zeta}}& \equiv &k^{\mathfrak{z}}\partial_{\mathfrak{z}} & \Rightarrow & k^{\mathfrak{z}} & = & {\rm i} \, \left(1 \, + \, \mathfrak{z}^2\right)& ; & \mbox{Disk-type, non-compact dilatation}
                                   \end{array} \right.
\end{equation}
This has important consequences on the structure of the momentum map leading to the $D$-type scalar potential and on the transformation properties of the fermions.
\par
Before proceeding further let us  stress once again that the choice of one or the other solution of the complex structure equation, that give to the foliations of $\Sigma$ into $\mathcal{G}$-orbits a different topology, depends on the global structure of the manifold $\Sigma$, whose metric we wrote in eq.(\ref{metraxia}). If we know a priori such a structure from an intrinsic definition of $\Sigma$ which arizes from other informations, than we know which complex structure is appropriate. Otherwise, choosing the complex structure amounts to the same as introducing one half of the missing information on the global structure of $\Sigma$, namely the range of the coordinate $B$. The other half is the range of the coordinate $U$. Actually as we shall  emphasize by means of the constant curvature examples that we are going to consider a criterion able to discriminate the relevant topologies is encoded in the asymptotic behavior of the function $\partial_C^2 J(C)$ for large and small values of its argument, namely in the center of the bulk and on the boundary.
\subsection{The Hodge bundle and Fayet-Iliopoulos Term}
Let us recall that relevant to supergravity is not only the K\"ahler structure of the surface $\Sigma$ rather the full-fledged geometry of the  Hodge bundle constructed over it.  Our surface $\Sigma$ is supposed to be  Hodge-K\"ahler and this implies that  there exists a line bundle $\mathcal{L} \rightarrow \Sigma$ whose Chern class coincides with the K\"ahler class, namely with the cohomology class of the K\"ahler two-form
$\mathrm{K} \, = \, {\rm i} \, g_{\mathfrak{z}\bar{\mathfrak{z}}} \,d\mathfrak{z} \, \wedge \, d\bar{\mathfrak{z}}$.
Explicitly we must have $c_1(L) \, = \,\left[\mathrm{K}\right]$, where the bracket denotes the cohomology class of the closed $p$-form embraced by it. The holomorphic sections of this line bundle are the possible superpotentials that encode the self interactions of the Wess-Zumino multiplet and its coupling to supergravity. The exponential of the K\"ahler potential is a fiber metric on the Hodge bundle:  for any holomorphic section $W(\mathfrak{z})$ of such a bundle we define an invariant  norm by means of the following position $
    {\left|| W \right||}^2 \, = \, \exp[\mathcal{K}] \, W(\mathfrak{z}) \, \overline{W}(\bar{\mathfrak{z}})$.
A fundamental object entering the construction of matter coupled supergravity is the logarithm of the superpotential norm $G(\mathfrak{z},\bar{\mathfrak{z}}) \, = \, \log \, || W||^2 \, = \, \mathcal{K} \, + \, \log \, W \, + \, \log \overline{W}$.
In the present paper, however, we do not consider this type of self-interactions and we put the superpotential to zero $W(z)=0$ so that we will just work with the K\"ahler potential $\mathcal{K}(\mathfrak{z},\bar{\mathfrak{z}})$.
Another fundamental ingredient in the matter coupling construction and in its gauging is provided by the prepotentials of the holomorphic Killing vectors. Following the discussion and the conventions of \cite{NoistandardN2}, if $k^\mathfrak{z} (\mathfrak{z})$, together with its complex conjugate $k^{\bar{\mathfrak{z}}} (\bar{\mathfrak{z}})$, is a holomorphic Killing vector, in the sense that the transformation:
\begin{equation}\label{transformazione}
    \mathfrak{z } \, \rightarrow \, z  \, + \, \epsilon  \, k^\mathfrak{z}  (\mathfrak{z})
\end{equation}
is an infinitesimal isometry of the K\"ahler metric for all choices of the small parameters $\epsilon $, then the prepotential of this Killing vector, which realizes the corresponding isometry   as a Lie-Poisson flux on the K\"ahler manifold, is the real function $\mathcal{P}(\mathfrak{z},\bar{\mathfrak{z}})= \mathcal{P}(\mathfrak{z},\bar{\mathfrak{z}})^\star$ defined by the following relations:
\begin{equation}\label{gospadi}
    k^\mathfrak{z} (\mathfrak{z}) \, = \, {\rm i} \, g^{\mathfrak{z}\bar{\mathfrak{z}}} \, \partial_{\bar{\mathfrak{z}}} \, \mathcal{P} \quad ; \quad k^{\bar{\mathfrak{z}}} (\bar{\mathfrak{z}}) \, = \, - \, {\rm i}
     \, g^{\mathfrak{z}\bar{\mathfrak{z}}}\, \partial_\mathfrak{z} \, \mathcal{P}
\end{equation}
In terms of the K\"ahler potential $\mathcal{K}$ function, supposedly invariant under the considered isometries, the Killing vector prepotential, satisfying the defining condition (\ref{gospadi}), is constructed through the following formula:
\begin{equation}\label{bozhemoi}
   \mathcal{P}\, = \,  \mathcal{P}_0 \, = \, {\rm i} \, \ft 12 \, \left(k^\mathfrak{z}\, \partial_\mathfrak{z} \,\mathcal{K} \,  - \, k^{\bar{\mathfrak{z}}}\,\partial_{\bar{\mathfrak{z}}}\,\mathcal{K} \right)
\end{equation}
It should be noted that the solution (\ref{bozhemoi}) of eq.(\ref{gospadi}) is defined up to an integration constant. Indeed setting:
\begin{equation}\label{ruffiano}
   \mathcal{P}\, = \,  \mathcal{P}_0 \, + \, \frac{\mathrm{q_f}}{g}
\end{equation}
where $\mathrm{q_f}$ is an arbitrary constant and $g$ is the gauge coupling constant equations (\ref{gospadi}) are still satisfied. It was first noted in \cite{Bagger:1982fn} that the above ambiguity is the mechanism behind the introduction of Fayet Iliopoulos terms into supersymmetric lagrangians \cite{Fayet:1974jb},\cite{Komargodski:2009pc}. The interpretation of Fayet Iliopoulos terms as constant shifts of the momentum maps was later extended to tri-holomorphic momentum maps and to the $\mathcal{N}=2$ theories in \cite{D'Auria:1990fj}. It should also be noted that the constant term $\frac{\mathrm{q_f}}{g}$ in the momentum can always be reabsorbed into $\mathcal{P}_0$, defined by eq. (\ref{bozhemoi}), introducing a new K\"ahler potential which differs from the first by a holomorphic K\"ahler transformation uneffective on the metric:
\begin{eqnarray}\label{kirghisistan}
    \tilde{\mathcal{K}}\left( \mathfrak{z}, \bar{\mathfrak{z}}\right)& = & {\mathcal{K}}\left( \mathfrak{z}, \bar{\mathfrak{z}}\right) \, + \, f(\mathfrak{z}) \, + \, \bar{f}(\bar{\mathfrak{z}})\nonumber\\
     {\rm i}\, k^\mathfrak{z} \,\partial_\mathfrak{z} \, f(\mathfrak{z}) &=& \frac{\mathrm{q_f}}{g}
\end{eqnarray}
Indeed note the second line of eq.(\ref{kirghisistan}) is a first order holomorphic differential equation that is always immediately solved by quadratures. Hence the appropriate function $f(\mathfrak{z})$ which produces the Fayet Iliopoulos term depends on the chosen Killing vector but the result on the momentum map is always the same: a constant shift.
\par
Upon gauging the isometry $B\to B+c$, the supergravity Lagrangian acquires a $D$-type potential proportional to the square of the momentum map $\mathcal{P}(\mathfrak{z},\bar{\mathfrak{z}})$:
\begin{equation}\label{YMpotus}
    V_{YM} \, \propto \, g^2 \, \left(\mathcal{P}(\mathfrak{z},\bar{\mathfrak{z}})\right)^2
\end{equation}
We shall come back to the discussion of such potentials. Before doing that we desire to illustrate some general features of the symmetries of the gauged supergravity lagrangian (particularly the fermionic sector) that heavily depend on the nature of the fundamental $B$-isometry.
\subsection{Sections of the Hodge bundle and the fermions}
The basic geometric mechanism that allows to gauge the global symmetries of $\mathcal{N}=1$ supergravity coupled to  Wess Zumino multiplets is the so named gauging of the composite connections. Let us recall such a notion.
The isometries of the K\"ahler metric  that take the infinitesimal form (\ref{transformazione}) extend to global symmetries of the full theory, including also the fermions, since all the items appearing in the lagrangian transform covariantly. From the geometrical point of view all fields are sections of the tangent bundle to the K\"ahler manifold and at the same time they are also sections of appropriate powers of the Hodge bundle. The subtle point is that under a holomorphic isometry:
$\mathfrak{z} \, \rightarrow\, \hat{\mathfrak{z}} \, = \,f(\mathfrak{z})$
the K\"ahler potential does not necessarily remain invariant rather it transform as follows:
\begin{equation}\label{ginocchio}
    \mathcal{K}\left(\hat{\mathfrak{z}},\hat{\bar{\mathfrak{z}}}\right) \, = \, \mathcal{K}\left({\mathfrak{z}},{\bar{\mathfrak{z}}}\right) \, + \, F(\mathfrak{z}) \, + \bar{F}(\bar{\mathfrak{z}})
\end{equation}
where $F(\mathfrak{z})$ is some holomorphic function associated with the considered transformation.  By definition a section
$\mathfrak{S}_p(\mathfrak{z})$ of weight $p$ of the Hodge-bundle transforms as follows
\begin{equation}\label{fontanafredda}
    \mathfrak{S}_p(\hat{\mathfrak{z}})\, = \, \mathfrak{S}_p(\mathfrak{z})\, \exp[- p \, F(\mathfrak{z})]
\end{equation}
 The fermion fields, namely the gravitino $\psi$, the chiralinos $\chi^\mathfrak{z}$, $\chi^{\bar{\mathfrak{z}}}$ and the gauginos $\lambda^\Lambda$ transform as sections of the Hodge bundle,  with half integer weights that we presently spell off.
\par
According to \cite{castadauriafre2}, we introduce the following notation for the chiral projections of the gravitino one-form $\psi$ and of the gaugino 0-forms $\lambda^\Lambda$ that are Majorana:
\begin{eqnarray}\label{gorgera}
    \psi & = & \psi_\bullet \, + \, \psi^\bullet\quad ; \quad \left \{ \begin{array}{rcl}
                                                                          \gamma_5 \, \psi_\bullet & =&\psi_\bullet \\
                                                                          \gamma_5 \, \psi^\bullet & =&-\, \psi^\bullet
                                                                        \end{array}
    \right.\nonumber\\
    \lambda & = & \lambda_\bullet \, + \, \lambda^{\bullet}\quad ; \quad \left\{ \begin{array}{rcl}
                                                                          \gamma_5 \, \lambda_\bullet & =&\lambda_\bullet \\
                                                                          \gamma_5 \, \lambda^{\bullet} & =&-\, \lambda^{\bullet}
                                                                        \end{array}\right.
\end{eqnarray}
while for the complex chiralino we simply have:
\begin{equation}\label{fermentilattici}
    \gamma_5 \, \chi^\mathfrak{z} \, = \, \chi^\mathfrak{z}\quad ; \quad \gamma_5 \, \chi^{\bar{\mathfrak{z}}} \, = \, - \,\chi^{\bar{\mathfrak{z}}}
\end{equation}
In what follows we just summarize and specialize to the minimal case of supergravity coupled to one vector multiplet $(1,\ft 12)$ and one WZ-multiplet $(\ft 12, 0^+, 0^-)$ what was described for the general case in \cite{primosashapietro}.
Having clarified the notation, the appropriate Hodge transformations for the fermions are:
\begin{equation}\label{pasticcaleone}
\begin{array}{ccccccc}
       \psi_\bullet & \to & \exp\left[{\rm i}\, \ft 12 \, F(z) \right]\, \psi_\bullet &\quad ; \quad&\psi^\bullet & \to & \exp\left[-\, {\rm i}\, \ft 12 \,  F(z) \right]\,  \psi^\bullet \\
       \lambda_\bullet & \to & \exp\left[{\rm i} \, \ft 12 \,  F(z) \right] \,\lambda_\bullet  &\quad ; \quad & \lambda^{\bullet}& \to & \exp\left[-\, {\rm i}\, \ft 12 \,  F(z) \right] \, \lambda^{\bullet} \\
       \chi^\mathfrak{z} & \to &\exp\left[- \, {\rm i}\, \ft 12 \,  F(z) \right] \, \chi^\mathfrak{z}  & ; & \chi^{\bar{\mathfrak{z}}} & \to  & \exp\left[{\rm i}\, \ft 12 \,  F(z) \right] \,\chi^{\bar{\mathfrak{z}}}
     \end{array}
\end{equation}
These transformations are compensated by the transformation of the Hodge bundle connection which is the following composite one-form:
\begin{equation}\label{HodgeKalloconno}
    Q \, \equiv \, {\rm i} \, \ft 12 \, \left( \partial_\mathfrak{z} \mathcal{K}\, d\mathfrak{z} \, - \, \partial_{\bar{\mathfrak{z}}} \mathcal{K} \, d\bar{\mathfrak{z}}\right)
\end{equation}
and enters the covariant derivatives of the fermions. For instance the gravitino and gaugino covariant derivatives are defined as follows:
\begin{eqnarray}\label{giannibeffa}
    \nabla \, \psi_\bullet & = & \mathcal{D}\psi_\bullet  \, + \, {\rm i} \, \ft 12 \, Q \, \wedge \, \psi_\bullet \quad ; \quad \mathcal{D}\psi_\bullet \, = \, d\psi_\bullet \, - \, \ft 14 \, \omega^{ab} \, \wedge \, \gamma_{ab} \, \psi_\bullet \nonumber\\
    \nabla \, \lambda_\bullet & = & \mathcal{D}\lambda_\bullet  \, + \, {\rm i} \, \ft 12 \, Q \,  \lambda_\bullet \quad\quad\,\, ; \quad \mathcal{D}\lambda_\bullet \, = \, d\lambda_\bullet \, - \, \ft 14 \, \omega^{ab} \,  \gamma_{ab} \, \lambda_\bullet
\end{eqnarray}
The gravitino one-form and the gaugino zero-forms have no indices along the tangent bundle of the K\"ahler manifold and therefore do not transform in the canonical bundle. On the other hand the chiralino carries a tangent space index and with respect to the canonical bundle it transforms  as a holomorphic vector. Correspondingly it enters the lagrangian covered by a covariant derivative of the form:
\begin{equation}
\nabla \,\chi^\mathfrak{z} \, \equiv\,  \mathcal{D}\chi^\mathfrak{z} \, + \, {\Gamma}^\mathfrak{z}_{\phantom{i}\mathfrak{z}} \chi^\mathfrak{z}
\,  - \, {\rm i} \ft 12 \, {Q} \, \chi^\mathfrak{z} \label{dchipredefi}
\end{equation}
In this way the isometries of the K\"ahler manifold are promoted to global symmetry of supergravity coupled, in the case under present consideration to just one vector multiplet.
\subsection{Gauging of the composite connections}The basic geometric mechanism that allows to gauge the above described global symmetries is the so named \textit{gauging of the composite connections}. Let us recall such a notion, according to the discussion of \cite{NoistandardN2} and \cite{primosashapietro}. In \cite{NoistandardN2}  the construction was applied to $\mathcal{N}=2$ supergravity so that the composite connections to be gauged were those emerging in Special K\"ahler Geometry. Here we focus on $\mathcal{N}=1$ supergravity and we just have Hodge-K\"ahler manifolds, yet the procedure is completely identical and it was already introduced in \cite{castadauriafre2}, but only for symmetries that are linearly realized on the scalars. In \cite{primosashapietro} it was smoothly generalized   to any type of holomorphic isometry, by means of the prepotential of the Killing vectors. In what follows we specialize the formulae of \cite{primosashapietro} to the minimal case here under discussion, where there is only one Wess-Zumino multiplet and only one isometry is gauged. The connections to be gauged are two: the Hodge-K\"ahler connection
(\ref{HodgeKalloconno})  and the Levi-Civita connection:
\begin{eqnarray}\label{levicivita}
    \Gamma^\mathfrak{z}_{\phantom{k}\mathfrak{z}} & = & \left\{\begin{array}{c}
                                             \mathfrak{z} \\
                                             \mathfrak{z} \,\mathfrak{z}
                                           \end{array}\right\}\,d\mathfrak{z} \, = \, g^{\mathfrak{z}\bar{\mathfrak{z}}} \, \partial_\mathfrak{z} \, g_{\bar{\mathfrak{z}}\mathfrak{z}} \, d\mathfrak{z}  \quad ; \quad
    \Gamma^{\bar{\mathfrak{z}}}_{\phantom{k}\bar{\mathfrak{z}}} \, = \, \mbox{Complex Conjugate of } \,   \Gamma^\mathfrak{z}_{\phantom{k}\mathfrak{z}}
\end{eqnarray}
We set :
\begin{equation}
                Q \,\to \, \widehat{Q} \, \equiv \, {\rm i} \, \ft 12 \, \left( \partial_\mathfrak{z} \mathcal{K}\, \nabla\mathfrak{z} \, - \, \partial_{\bar{\mathfrak{z}}} \mathcal{K} \, \nabla\bar{\mathfrak{z}}\right) \quad ; \quad
                \Gamma^\mathfrak{z}_{\phantom{k}\mathfrak{z}} \,\to \, \widehat{\Gamma}^\mathfrak{z}_{\phantom{k}\mathfrak{z}}  \, = \, \left\{\begin{array}{c}
                                             \mathfrak{z} \\
                                            \mathfrak{z} \,\mathfrak{z}
                                           \end{array}\right\}\,\nabla\mathfrak{z}
              \end{equation}
where
\begin{equation}\label{cirimella}
    \nabla \mathfrak{z} \, = \, d\mathfrak{z} \, + \, g\, \mathcal{A} \, k^\mathfrak{z}(\mathfrak{z})
\end{equation}
is the covariant derivative of the complex scalar field, $g$ being the gauge coupling constant, $\mathcal{A} \, = \, \mathcal{A}_\mu \,dx^\mu$ the gauge field one-form and $k^\mathfrak{z}(\mathfrak{z})$ the Killing vector.
It follows from the various identities presented above that:
\begin{equation}
% \nonumber to remove numbering (before each equation)
  \widehat{Q}\,= \, Q \, + \, g \, \mathcal{A} \, \mathcal{P} \quad ;\quad
  \widehat{\Gamma}^\mathfrak{z}_{\phantom{i}\mathfrak{z}}  \, = \, {\Gamma}^\mathfrak{z}_{\phantom{i}\mathfrak{z}} \, + \, g \,\mathcal{A}  \, \partial_{\mathfrak{z}}\,k^\mathfrak{z}(\mathfrak{z}) \label{bertoldino}
\end{equation}
\par
\section{General features and symmetries of the minimal supergravity inflationary model.}
\label{gensec}
Without entering into the details of any specific model there is a number of features  that immediately follow from the formulae presented above, which can be discussed in general terms and significantly distinguish the two cases of gauging either a rotation or a shift symmetry as basic mechanism for the generation of an inflaton potential. These general properties are in our opinion more important and fundamental than the specific form of the inflaton potential obtained from the gauging.
\subsection{Compact $\mathrm{U(1)}$ case}
If the fundamental isometry of $\Sigma$ is a compact $\mathrm{U(1)}$, the Killing vector is given by the first line of eq.(\ref{familione}) and we have:
\begin{equation}\label{carezza1}
    \partial_\mathfrak{z} \,k^\mathfrak{z} \, = \, {\rm i}
\end{equation}
so that from equation (\ref{bertoldino}) we obtain:
\begin{equation}\label{fonzi1}
    \widehat{\Gamma}^\mathfrak{z}_{\phantom{i}\mathfrak{z}}  \, = \, {\Gamma}^\mathfrak{z}_{\phantom{i}\mathfrak{z}} \, + \,{\rm i} \, g \, \mathcal{A}  \,
\end{equation}
Furthermore, given a $\mathrm{U(1)}$-invariant K\"ahler potential $\tilde{\mathcal{K}}(\mathfrak{z},\mathfrak{z})$, namely such  that:
\begin{equation}\label{invariatur}
   \left( \mathfrak{z} \, \partial_\mathfrak{z} \, - \, \bar{\mathfrak{z}} \, \partial_{\bar{\mathfrak{z}}} \right) \tilde{\mathcal{K}}(\mathfrak{z},\mathfrak{z}) \, = \, 0
\end{equation}
and the form of the Killing vector mentioned in the first line of eq.(\ref{familione}), the solution  of eq.(\ref{kirghisistan}) is the following one:
\begin{equation}\label{superpottuslano}
    f(\mathfrak{z}) \, = \,  \log \,\mathfrak{z}^{{q}_{\mathrm{f}}/g}
\end{equation}
where ${q}_{\mathrm{f}}/g$ corresponds to the Fayet Iliopoulus charge introduced in eq.(\ref{ruffiano}).  Setting ${\mathcal{K}} \, = \, \tilde{\mathcal{K}} \, + \, f(\mathfrak{z}) + \bar{f}(\bar{\mathfrak{z}})$  we obtain
\begin{equation}\label{giuleppo}
    \hat{\mathcal{P}}(\mathfrak{z},\mathfrak{z}) \, = \,  \, \tilde{\mathcal{P}}(\mathfrak{z},\mathfrak{z})  \, + \, \frac{{q}_{\mathrm{f}}}{g}
\end{equation}
Note that if $\tilde{\mathcal{K}}$ is $\mathrm{U(1)}$ invariant the same is true of ${\mathcal{K}}$.
As already stressed the constant shift ${q}_{\mathrm{f}}/g $ of the momentum map  has no effect on the K\"ahler metric and, consequently,  on the kinetic terms of the scalar fields. Actually it survives at vanishing scalar fields and    it exists even if we completely suppress the Wess-Zumino multiplet.
\par
As a consequence of the above formulae the covariant derivatives of the fermions entering the minimal supergravity lagrangian are the following ones:
\begin{eqnarray}
    \nabla \, \psi_\bullet & = & \widetilde{\nabla}\psi_\bullet  \, - \, {\rm i} \,\ft 12 \, {q}_{\mathrm{f}}\, \mathcal{A} \,\wedge \, \psi_\bullet \quad \quad \, ; \quad\widetilde{\nabla}\psi_\bullet = \, \mathcal{D}\psi_\bullet \, + \, {\rm i} \,\ft 12\, Q\, \wedge \, \psi_\bullet \, + \, {\rm i} \,\ft 12  \, g \,\tilde{\mathcal{P}}\, \mathcal{A}\,  \wedge \,  \psi_\bullet \nonumber\\
    \nabla \, \lambda_\bullet & = & \widetilde{\nabla}\lambda_\bullet  \, - \, {\rm i} \,\ft 12 \, {q}_{\mathrm{f}}\, \mathcal{A} \,  \lambda_\bullet \quad \quad \quad \,\,\, ; \quad\widetilde{\nabla}\lambda_\bullet = \, \mathcal{D}\lambda_\bullet \, + \, {\rm i} \,\ft 12\, Q\, \lambda_\bullet \, + \, {\rm i} \,\ft 12  \, g \, \mathcal{A}\, \tilde{\mathcal{P}} \,  \lambda_\bullet \nonumber\\
\nabla \,\chi^\mathfrak{z} & = &  \tilde{\nabla}\chi^\mathfrak{z} \, +  {\rm i} \,\ft 12 \, \left({q}_{\mathrm{f}}+ 2\,g\right)\, \mathcal{A}\, \chi^\mathfrak{z} \quad ; \quad   \widetilde{\nabla}\chi^\mathfrak{z} \, =  \mathcal{D}\chi^\mathfrak{z} \, +\, {\Gamma}^\mathfrak{z}_{\phantom{i}\mathfrak{z}} \chi^\mathfrak{z}
\,  - \, {\rm i} \ft 12 \, {Q} \, \chi^\mathfrak{z} \, - \, {\rm i} \,\ft 12  \, g  \tilde{\mathcal{P}} \,  \mathcal{A}\,\chi^\mathfrak{z} \label{giannibuffone}
\end{eqnarray}
At the same time the covariant derivative of the complex scalar field is:
\begin{equation}\label{paratto}
    \nabla \mathfrak{z} \,=\, d\mathfrak{z} \, + \,{\rm i} \, g \, \mathcal{A} \, \mathfrak{z}
\end{equation}
Equations (\ref{giannibuffone}) and (\ref{paratto}) are sufficient to draw the main conclusions concerning the symmetries of the supergravity lagrangian.
\par
\begin{description}
  \item[a)] There is one chiral global $\mathrm{U_R(1)}$ symmetry (the $R$-symmetry)
  with respect to which the gravitino and the gaugino have charge\footnote{The spelled out charges are those of the holomorphic-chiral fields (left-handed the fermions, holomorphic the scalar). The charges of the antiholomorphic-antichiral fields (right-handed the fermions, antiholomorphic the scalar) are just the opposite ones.} $q_R(\psi)\, = \,q_R(\lambda)  \, = \, \ft 12$, the chiralino has charge $q_R(\chi) \, = \, -\ft 12$ and the scalar has charge $q_R(\mathfrak{z}) \, = \, 0$. Note the $R$-symmetry charges are the same as the K\"ahler weights of the corresponding fields under change of trivializations in the Hodge bundle.
  \item[b)] There is another chiral global  $\mathrm{U_B(1)}$ symmetry with respect to which the gravitino and the gaugino have charge $q_{B}(\psi)\,=\,q_{B}(\lambda)\, = \, 0$,  the chiralino has charge $q_{B}(\chi) \, = \, 1$ and the scalar has charge $q_{B}(\mathfrak{z}) \, = \, 1$.
  \item[c)] When we gauge the model in the absence of a Fayet Iliopoulos term the actual gauge algebra is just:
  \begin{equation}\label{furbo}
    \mathfrak{g}_{gauge} \, = \, g \, \uu_B(1)
  \end{equation}
  \item[d)] When we gauge the model in the presence of a Fayet Iliopoulos term the actual gauge group is just:
  \begin{equation}\label{furbo}
    \mathfrak{g}_{gauge} \, = \, - \, {q}_{\mathrm{f}} \, \uu_R(1) \, \oplus \, \, g \, \uu_B(1)
  \end{equation}
  \item[e)] If we put $g\,=\, 0$ and we just remove the Wess-Zumino multiplet by setting  $\mathfrak{z}\, =\, \chi^\mathfrak{z} \, = \,0$  we can nonetheless preserve a non vanishing  Fayet Iliopoulos charge ${q}_{\mathrm{f}} \, \ne \, 0$. This means that the gauge field $\mathcal{A}$ is utilized to gauge $R$-symmetry and this produces a positive cosmological constant $\Lambda \, = \, {q}^2_{\mathrm{f}} $ which yields a de-Sitter vacuum where supersymmetry is broken. This is the model constructed by Freedman in \cite{Freedman:1976uk}.
  \end{description}
\subsection{Non compact shift-symmetry}
If the fundamental isometry of $\Sigma$ is a non-compact translation symmetry $\mathbb{R}$, the Killing vector is given by the second line of eq.(\ref{familione}) and we have:
\begin{equation}\label{carezza2}
    \partial_\mathfrak{z} \,k^\mathfrak{z} \, = \, 0
\end{equation}
so that from equation (\ref{bertoldino}) we obtain:
\begin{equation}\label{fonzi2}
    \widehat{\Gamma}^\mathfrak{z}_{\phantom{i}\mathfrak{z}}  \, = \, {\Gamma}^\mathfrak{z}_{\phantom{i}\mathfrak{z}}
\end{equation}
Furthermore, given a shift-invariant K\"ahler potential $\tilde{\mathcal{K}}(\mathfrak{z},\mathfrak{z})$, namely such  that:
\begin{equation}\label{invariatur}
   \left( \partial_\mathfrak{z} \, + \, \partial_{\bar{\mathfrak{z}}} \right) \tilde{\mathcal{K}}(\mathfrak{z},\mathfrak{z}) \, = \, 0
\end{equation}
the solution of eq.(\ref{kirghisistan}) for the K\"ahler gauge transformation producing a Fayet Iliopoulos charge  is the following one:
\begin{equation}\label{superpottuslano}
    f(\mathfrak{z}) \, = \,  {\rm i} \, \frac{q_{\mathrm{f}}}{g} \mathfrak{z}
\end{equation}
leading to
\begin{equation}\label{siggio}
    \mathcal{K}(\mathfrak{z},\mathfrak{z}) \, = \, \tilde{\mathcal{\mathcal{K}}}(\mathfrak{z},\mathfrak{z}) \, - \, 2 \, \frac{q_{\mathrm{f}}}{g} \, \mathrm{Im} \,\mathfrak{z}
\end{equation}
and to a momentum map with the same structure as in eq.(\ref{giuleppo}).
As a consequence of the such formulae, the covariant derivatives of the fermions entering the minimal supergravity lagrangian are now the following ones to be compared with eq.s(\ref{giannibuffone}):
\begin{eqnarray}
    \nabla \, \psi_\bullet & = & \widetilde{\nabla}\psi_\bullet  \, - \, {\rm i} \,\ft 12 \, {q}_{\mathrm{f}}\, \mathcal{A} \,\wedge \, \psi_\bullet \quad \quad \, ; \quad\widetilde{\nabla}\psi_\bullet = \, \mathcal{D}\psi_\bullet \, + \, {\rm i} \,\ft 12\, Q\, \wedge \, \psi_\bullet \, + \, {\rm i} \,\ft 12  \, g \,\tilde{\mathcal{P}}\, \mathcal{A}\,  \wedge \,  \psi_\bullet \nonumber\\
    \nabla \, \lambda_\bullet & = & \widetilde{\nabla}\lambda_\bullet  \, - \, {\rm i} \,\ft 12 \, {q}_{\mathrm{f}}\, \mathcal{A} \,  \lambda_\bullet \quad \quad \quad \,\,\, ; \quad\widetilde{\nabla}\lambda_\bullet = \, \mathcal{D}\lambda_\bullet \, + \, {\rm i} \,\ft 12\, Q\, \lambda_\bullet \, + \, {\rm i} \,\ft 12  \, g \, \mathcal{A}\, \tilde{\mathcal{P}} \,  \lambda_\bullet \nonumber\\
\nabla \,\chi^\mathfrak{z} & = &  \tilde{\nabla}\chi^\mathfrak{z} \, +  {\rm i} \,\ft 12 \, {q}_{\mathrm{f}}\, \mathcal{A}\, \chi^\mathfrak{z} \quad \quad \quad \, \, \, \, ; \quad   \widetilde{\nabla}\chi^\mathfrak{z} \, =  \mathcal{D}\chi^\mathfrak{z} \, +\, {\Gamma}^\mathfrak{z}_{\phantom{i}\mathfrak{z}} \chi^\mathfrak{z}
\,  - \, {\rm i} \ft 12 \, {Q} \, \chi^\mathfrak{z} \, - \, {\rm i} \,\ft 12  \, g  \tilde{\mathcal{P}} \,  \mathcal{A}\,\chi^\mathfrak{z} \label{giannibuffone2}
\end{eqnarray}
and the covariant derivative of the complex scalar field is:
\begin{equation}\label{paratto}
    \nabla \mathfrak{z} \,=\, d\mathfrak{z} \, + \, g \, \mathcal{A} \, \Rightarrow \, \left\{\begin{array}{rcl}
                                                                                                 \nabla C& = & dC \\
                                                                                                \nabla B & = & dB \, + \, g \, \mathcal{A}
                                                                                              \end{array}
    \right.
\end{equation}
It follows that:
\begin{description}
  \item[a)] Just as before there is  one chiral global $\mathrm{U_R(1)}$ symmetry (the $R$-symmetry)
  with respect to which the gravitino and the gaugino have charge $q_R(\psi)\, = \,q_R(\lambda)  \, = \, \ft 12$, the chiralino has charge $q_R(\chi) \, = \, -\ft 12$ and the scalar has charge $q_R(\mathfrak{z}) \, = \, 0$. Note the $R$-symmetry charges are the same as the K\"ahler weights of the corresponding fields under change of trivializations in the Hodge bundle.
  \item[b)] The second chiral  global symmetry $\mathrm{U_B(1)}$ which is present in the compact case, here is absent.
  \item[d)] In the whole lagrangian the $B$-field appears only under derivatives.
  \item[c)] When we gauge the model in the absence of a Fayet Iliopoulos term the actual gauge algebra is just:
  \begin{equation}\label{furbo}
    \mathfrak{g}_{gauge} \, = \, g \, \mathbb{R}
  \end{equation}
  all the fermions are neutral under such an algebra and the gauge field $\mathcal{A}$ appears only in the combination
  $\mathcal{A} \, + \, \frac{1}{g} dB$ that can be renamed $\hat{\mathcal{A}}$ and describes a massive vector field. The massive vector field $\hat{\mathcal{A}}$, the inflaton scalar $C$ and the two fermions $\lambda$, $\chi$ make up the field content of a massive vector multiplet with $4$-bosonic degrees of freedom $\oplus$ $4$-fermionic degrees of freedom.
  \item[d)] When we gauge the model in the presence of a Fayet Iliopoulos term the actual gauge group is just:
  \begin{equation}\label{furbo}
    \mathfrak{g}_{gauge} \, = \, - \, {q}_{\mathrm{f}} \, \uu_R(1) \, \oplus \, \, g \, \mathbb{R}
  \end{equation}
  \item[d)] As in the previous compact case, if we put $g\,=\, 0$ and we just remove the Wess-Zumino multiplet by setting  $\mathfrak{z}\, =\, \chi^\mathfrak{z} \, = \,0$  we can nonetheless preserve a non vanishing  Fayet Iliopoulos charge ${q}_{\mathrm{f}} \, \ne \, 0$. Also in this case the gauge field $\mathcal{A}$ is utilized to gauge $R$-symmetry and this produces a positive cosmological constant $\Lambda \, = \, {q}^2_{\mathrm{f}} $ which yields a de-Sitter vacuum where supersymmetry is broken. Actually, once the WZ-multiplet is removed, the distinction between the compact and the shift-symmetry case is removed and we just have the already mentioned gauging of $R$-symmetry. Once again this is the model constructed by Freedman in \cite{Freedman:1976uk}.
  \end{description}
\section{Constant Curvature Models}
%{The Hodge bundle and Fayet-Iliopoulos term}
Having established the above general facts, in the present section we consider explicit examples classified according to the curvature of the K\"ahler  surface $\Sigma$.
\par
First we consider  flat models where,  written, in a standard complex coordinate $\mathfrak{z}$, the K\"ahler metric is $ds^2 \, \propto \, d\bar{\mathfrak{z}} \, d\mathfrak{z}$. Next we  consider constant (negative) curvature models, where, written in a disk-type complex coordinate $\mathfrak{z} \, = \, \zeta$, the  K\"ahler metric is $ds^2 \, \propto \, \left(1 \, - \, |\zeta|^2\right )^{-2} \, d\bar{\zeta} \, d\zeta$. We show that
the \textit{a priori} knowledge of the form of the metric in a standard complex coordinate is precisely what allows to determine the appropriate solution of the complex structure equations and, as a by product, to determine the global structure of the isometry group generated by the Killling vector $\vec{k}_B$. In want of this knowledge one has to resort to the criterion
of the asymptotic behavior of the function $\partial_C^2 \, J(C)$ in order to discriminate between the possible topologies. We analyze from this point of view the constant curvature models in order to verify the established criteria.
\par
\subsection{The curvature and the K\"ahler potential}
The curvature of an axial (shift) symmetric K\"ahler manifold can be written in two different ways in terms of the canonical coordinate $\phi$ or the VP coordinate $C$. In terms of the VP coordinate $C$ we have the following formula:
\begin{eqnarray}\label{curvatta}
  R & = &  R(C) \, = \, \frac{J^{'''}(C)^2-J''(C) J^{''''}(C)}{2 \, J''(C)^3}
    \, = \, -\, \ft 12 \, \partial_C^2 \, \log \left[ \partial_C^2 J(C) \right ] \, \frac{1}{\partial_C^2 J(C)}
\end{eqnarray}
which can be derived from the standard structural equations of the manifold:
\begin{eqnarray}
% \nonumber to remove numbering (before each equation)
0 &=&  \mathrm{d}E^1 \, + \, \omega \, \wedge \, E^2 \nonumber\\
0 &=&  \mathrm{d}E^2 \, - \, \omega \, \wedge \, E^1  \nonumber\\
\mathfrak{R} & \equiv & \mathrm{d}\omega \, \equiv \, R \, E^1 \, \wedge \, E^2 \label{garducci}
\end{eqnarray}
by inserting into them the appropriate form of the zweibein:
\begin{equation}\label{zweibeinOne}
    E^1 \, = \, \ft 12 \, \sqrt{J''(C)} \, dC \quad ; \quad E^2 \, = \, \ft 12 \, \sqrt{J''(C)} \, dB \quad \Rightarrow \quad ds^2 \, = \, \ft 14 \, J''(C) \left( dC^2 \, + \, dB^2\right)
\end{equation}
Alternatively we can write the curvature in terms of the momentum map $\mathcal{P}(\phi)$ or of the D-type potential
 $V(\phi) \, \propto \, \mathcal{P}^2(\phi)$ if we use the canonical coordinate $\phi$ and the corresponding appropriate zweibein:
\begin{equation}\label{zweibeinTwo}
    E^1 \, = \, \ft 12  \, d\phi \quad ; \quad E^2 \, = \, \ft 12 \, \mathcal{P}^\prime(\phi) \, dB \quad \Rightarrow \quad ds^2 \, = \, \ft 14 \, \left( d\phi^2 \, + \, \left(\mathcal{P}^\prime(\phi) \right)^2 \, dB^2\right)
\end{equation}
Upon insertion of eq.s (\ref{zweibeinTwo}) into (\ref{garducci}) we get:
\begin{eqnarray}\label{giunone}
     R(\phi) & = & - 4 \, \frac{\mathcal{P}^{\prime\prime\prime}(\phi)}{\mathcal{P}^{\prime}(\phi)} \,= \, -\, 4 \, \left ( \frac{V^{\prime\prime\prime}}{V^{\prime}} \, - \, \ft 32 \, \frac{V^{\prime\prime}}{V} \, - \, \ft  34 \, \left( \frac{V^\prime}{V}\right)^2 \right)
\end{eqnarray}
Finally let us compare the above definition (\ref{curvatta}) of the curvature with that utilized in curved index formalism. For instance, if we consider the Disk-type complex structure and we set
\begin{equation}\label{rostov}
    R \, = \, - \, \ft 12 \, g^{t\bar{t}} \, \partial_t \, \partial_{\bar{t}} \, \log \, g_{t\bar{t}}
\end{equation}
we just reproduce the result (\ref{curvatta}), since $ g_{t\bar{t}} \, = \, \ft 14 \, J^{\prime\prime}(C)$ and $\partial_t \, \partial_{\bar{t}} \, \simeq \, \ft 14 \, \partial_C^2$.
Eq.(\ref{giunone}) was first derived in \cite{primosashapietro} \footnote{The curvature $R$ defined as the component of $d\omega$ in the basis $E^1 \wedge E^2$ differs by a factor $\ft 12$ with respect to the curvature defined in standard curved index tensor calculus. This difference sums up in explicit calculations with the difference in normalization of the scalar field $\phi$ (see Friedman equations (\ref{fridmano})).}.
\subsection{Flat models}
The above formulae for the curvature easily allow an analysis of the simplest possible supergravity models, namely those based on a flat K\"ahler manifold where $R=0$. It is quite instructive to implement the vanishing curvature condition in both formulations (\ref{curvatta}) and (\ref{giunone}).
\subsubsection{Canonical coordinate representation}
If we start from eq.(\ref{giunone}), we see that the most general solution of the vanishing curvature condition is:
\begin{equation}\label{firtullo}
    \mathcal{P}(\phi) \, = \, a_0 \, + \, a_1 \, \phi \, + \, \ft 12 \, a_2 \, \phi^2
\end{equation}
where $a_{0,1,2}$ are real constants. By means of the shift  $\phi \, \to \, \phi \, - \, \frac{a_1}{a_2}$, which does not alter the canonical kinetic term of $\phi$, we can always suppress the linear term $a_1 \, = \, 0$ and we are left with:
\begin{equation}\label{fardallo}
    \mathcal{P}(\phi) \, = \, \left( a_0 \, + \ft 12 \,  a_2 \, \phi^2 \right) \quad \Rightarrow \quad V(\phi) \, \propto \, \left(a_0 \, + \, \ft 12 \, a_2 \, \phi^2 \right)^2 \, = \, \underbrace{M^4 \left[ \left(\frac{\phi}{\phi_0} \right)^2 \, \pm \, 1 \right]}_{\mbox{ $M^4 \, = \, a_0^2$ ; $\phi_0^2 \, = \, \left|2 \, \frac{a_0}{a_2}\right|$}}
\end{equation}
where the choice of the sign $\pm$ depends on whether $\frac{a_0}{a_2} > 0$ or $\frac{a_0}{a_2} < 0$. In the second case the obtained potential is the Higgs type of quartic potential that in the classification of inflationary potentials presented in \cite{Encyclopaedia} has the name DWI (see table 1 of the quoted reference).
\paragraph{The general case $a_2\ne0$}
Eq.(\ref{fardallo}) shows that when both  $a_2$ does not vanish  the Higgs type of quartic potential can be incorporated into supergravity based on a flat K\"ahler manifold. Applying eq.(\ref{sodoma}) to the present case we obtain the following relation between the VP coordinate and the canonical coordinate
\begin{equation}\label{flattaOne}
  C(\phi) \, \equiv \,  \int \frac{d\phi}{\mathcal{P}^\prime(\phi)} \, = \, \frac{\log (\phi )}{a_2}  \quad \Rightarrow \quad \phi \, = \, e^{C a_2}
\end{equation}
while the K\"ahler potential is given by:
\begin{eqnarray}
  J(\phi) &=& \int \frac{\mathcal{P}(\phi)}{\mathcal{P}^\prime(\phi)} \, = \, \frac{\phi ^2}{4}+\frac{\log
   (\phi ) a_0}{a_2} \, + \, \mbox{const}\nonumber\\
  \null &\Downarrow& \null \nonumber \\
  J(C) &=& C a_0+\frac{1}{4} e^{2 C a_2}+ \, \mbox{const} \label{cellerusOne}
\end{eqnarray}
Correspondingly the metric takes the following form:
\begin{equation}\label{metrozzata}
    ds^2 \, = \, \frac{a_2^2}{4} \, \left(\mathrm{d}B^2+\mathrm{d}C^2\right) e^{2 \, a_2 \, C}
\end{equation}
and it is turned into the standard form of the flat K\"ahler metric:
\begin{equation}\label{frollino}
    ds^2_{flat} \, \equiv \, \frac{1}{4} \, d\mathfrak{z} \, d\bar{\mathfrak{z}} \quad \Leftrightarrow \quad \mathcal{K}\left (\mathfrak{z}, \bar{\mathfrak{z}}\right) \, = \, \frac{1}{4} \, \left(  \mathfrak{z} \, \bar{\mathfrak{z}} \, - \, \frac{2\, a_0}{a_2} \, \log\left[\mathfrak{z} \, \bar{\mathfrak{z}}\right] \right)
\end{equation}
by the identification:
\begin{equation}\label{malicus}
    \mathfrak{z} \, = \, \exp \left [ a_2 \left(C \, + \, {\rm i} B\right)\right]
\end{equation}
It follows that when $a_2 \,\ne \,0$, the proper interpretation of the symmetry $B \, \rightarrow \, B \, + \, c$, which is gauged in order to produce the potential, is that of a \textbf{compact rotation}. Furthermore the parameter $\frac{ a_0}{2 \,a_2}$ plays the role of a Fayet-Iliopoulos charge according to the discussion of section \ref{gensec}.
%%%%%%%%%%%%%%%%%%%%%%%%%%%
\begin{figure}[!hbt]
\begin{center}
\iffigs
\includegraphics[height=70mm]{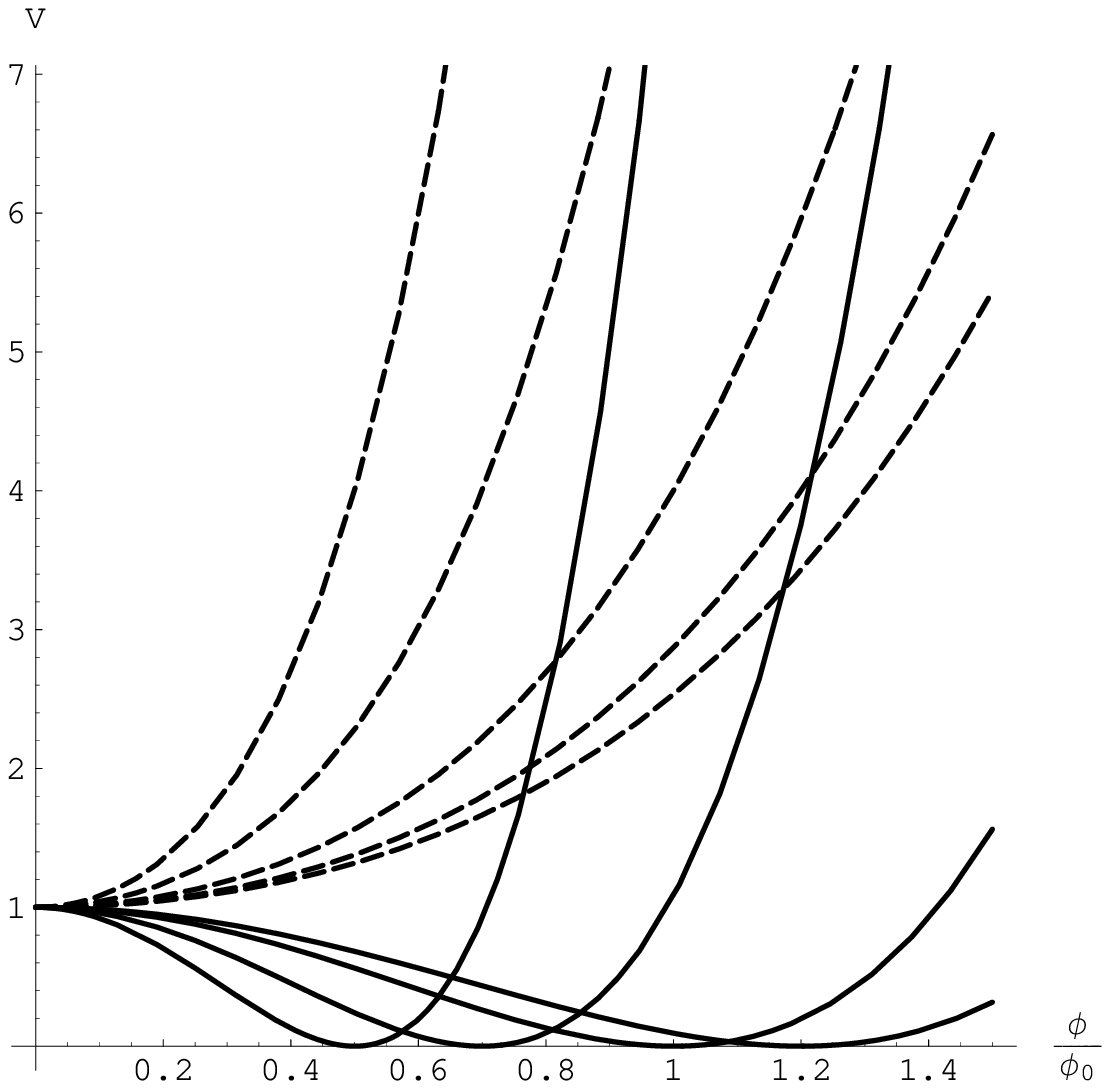}
\includegraphics[height=70mm]{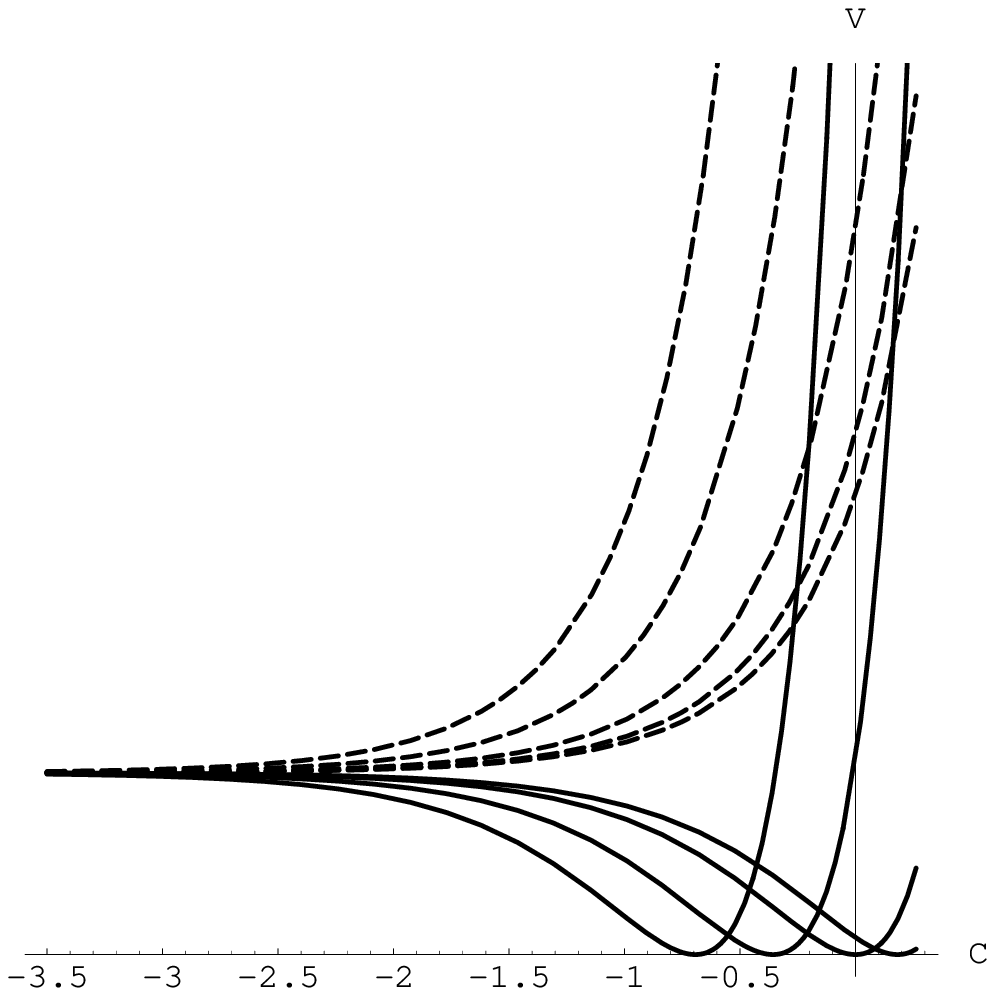}
\else
\end{center}
 \fi
\caption{\it  In this figure we show the plots of the inflaton potential obtained from minimal supergravity with a flat K\"ahler metric in the case of the gauging of a $\mathrm{U(1)}$ symmetry. The plots are presented in terms of the canonical variable $\phi$ (picture on the left) and in terms of the VP coordinate $C$ (picture on the right). The various curves correspond to different values of the parameter $\phi_0 \,= \, \sqrt{2\left|\frac{a_0}{a_2}\right|}$. The solid curves correspond to the cases where $\frac{a_0}{a_2} <0$, while the dashed curves correspond to the cases $\frac{a_0}{a_2} > 0$. In the first case the plots in terms of $\phi$ are one half of the familiar mexican hat shaped Higgs potentials. The restriction $\phi \ge 0$ are due to the relation $\phi \, \propto \, \exp[C]$.}
\label{piattosiOne}
 \iffigs
 \hskip 1cm \unitlength=1.1mm
 \end{center}
  \fi
\end{figure}
The plots of these type of potentials are displayed in fig.\ref{piattosiOne}.
\paragraph{The case $a_2 = 0$, $a_1 \ne 0$.} As it is evident from the above formulae the limit $a_2 \, \to \, 0$ is singular and the case $a_2 \, = \, 0$ has to be treated separately. In the case that the momentum map is linear in the canonical coordinate, eq.s (\ref{flattaOne}) and (\ref{cellerusOne}) are replaced by:
\begin{equation}\label{flattaOne}
  C(\phi) \, \equiv \,  \int \frac{d\phi}{\mathcal{P}^\prime(\phi)} \, = \, \frac{\phi }{a_1} \, - \, \frac{\beta}{a_1^2} \quad \Rightarrow \quad \phi \, = \,   \frac{C \, a_1^2\, +\, \beta }{a_1}
\end{equation}
and:
\begin{eqnarray}
  J(\phi) &=& \int \frac{\mathcal{P}(\phi)}{\mathcal{P}^\prime(\phi)} \, = \,  \ft 12 \,\phi^2  + \, \mbox{const}\nonumber\\
  \null &\Downarrow& \null \nonumber \\
  J(C) &=&  \frac{1}{2} \, a_1^2 \, C^2 \, + \, \beta \, C \, + \, \mbox{const}^\prime \label{cellerusTwo}
\end{eqnarray}
where $- \, \frac{\beta}{a_1^2}$ is an arbitrary integration constant. In this case the metric is:
\begin{equation}\label{buffido}
    ds^2 \, = \, a_1^2 \, \left(\mathrm{d}B^2+\mathrm{d}C^2\right)
\end{equation}
which is turned into the standard form of the flat metric:
\begin{equation}\label{frollinoTwo}
    ds^2_{flat} \, \equiv \, a_1^2 \, d\mathfrak{z}\, d\bar{\mathfrak{z}} \quad \Leftrightarrow \quad \mathcal{K}\left (\mathfrak{z} ,  \bar{\mathfrak{z}}\right) \, = \, - \, \ft 18 \, a_1^2\, \left(\mathfrak{z} \, - \, \bar{\mathfrak{z}}\right)^2 \, - \, {\rm i} \, \ft 12 \, \beta \, \left(\mathfrak{z} \, - \, \bar{\mathfrak{z}}\right)
\end{equation}
by the identification:
\begin{equation}\label{malicus2}
    \mathfrak{z} \, = \, {\rm i} \, C \, +  B
\end{equation}
It follows that when $a_2 \,= \,0$, the proper interpretation of the symmetry $B \, \rightarrow \, B \, + \, c$ which is gauged in order to produce the potential is that of a \textbf{non compact translation}, namely of a shift symmetry. The integration constant $\beta$ plays now the role of Fayet Iliopoulos gauging constant. From the point of view of the scalar potential this case corresponds to a pure mass term:
\begin{equation}\label{grullino}
    \mathcal{P}(\phi) \, = \, a_0 \, + \, a_1 \, \phi \quad \Rightarrow \quad V(\phi) \, = \, \left(a_0 \, + \, a_1\, \phi\right)^2
\end{equation}
%%%%%%%%%%%%%%%%
\begin{figure}[!hbt]
\begin{center}
\iffigs
\includegraphics[height=70mm]{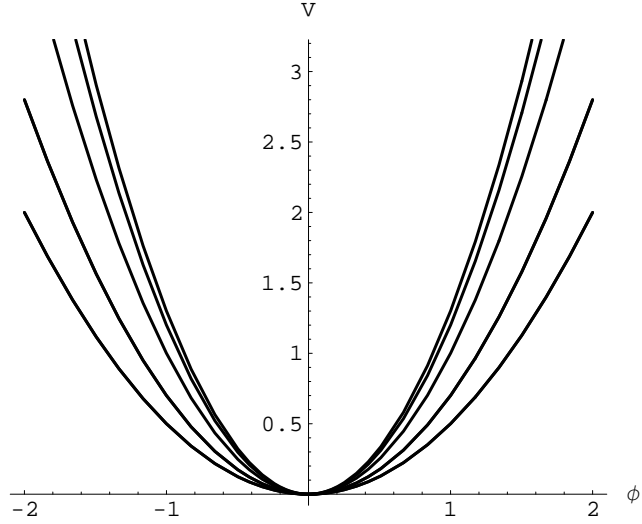}
\else
\end{center}
 \fi
\caption{\it  In this figure we show the plots of the inflaton potential obtained from minimal supergravity with a flat K\"ahler metric in the case of the gauging of a shift symmetry. In this case the canonical variable $\phi$ and the VP coordinate $C$ coincide. The various curves correspond to different values of the coefficient $a_1$.  The coefficient $a_0$ can always be set to zero by means of a constant shift of $\phi$ which does not alter the metric. In all cases the potentials are quadratic and in the language of inflationary models correspond to \textbf{chaotic inflation}}
\label{piattosiTwo}
 \iffigs
 \hskip 1cm \unitlength=1.1mm
 \end{center}
  \fi
\end{figure}
%%%%%%%%%%%%%%%%%%%%%%%%%%%
The plot of these type of potentials are displayed in fig.\ref{piattosiTwo}.
\subsubsection{Retrieving the same result in the VP coordinate representation} If we start from equation (\ref{curvatta}), by imposing the zero curvature condition we obtain that $\log \left(\partial^2_C J\right)$ should be linear in $C$, namely
\begin{equation}\label{costallico}
   \log \left(\partial^2_C J\right) \, = \, 2 \, a_2 \, C \, + \, \log \left(a_2^2\right) \quad \Rightarrow \quad \,   \partial^2_C J \, = \, a_2^2 \, \exp\left[ 2 \, a_2 \, C\right] \quad \Rightarrow \quad J \, = \, \frac{1}{4} \, \exp\left[ 2 \, a_2 \, C\right] \, + \, a_0 \, C \, + \, \mbox{const}
\end{equation}
where $a_0$ is the name given to the integration constant in the second order differential equation displayed above. This makes immediate contact with the result obtained from the momentum map approach. Note that the constant term in the solution of the differential equation for $\log \left(\partial^2_C J\right)$  simply amounts to the  rescaling of the K\"ahler potential and, hence, of the K\"ahler metric by an overall constant. The choice of that constant equal to $\log \left(a_2^2\right)$ simply fixes the standard normalization of the scalar field kinetic term.
\subsubsection{Asymptotic expansions of the function $\partial_C^2 J(C)$}
Let us now discuss the behavior of the function $\partial_C^2 J(C)$ which determines the K\"ahler metric in real variables in the two instances of flat models discussed above.
\paragraph{The flat $\mathrm{U(1)}$ model.} For $C\to 0$ the metric coefficient  $\partial_C^2 J(C) \, = \, a_2^2 \, \exp \,\left [ 2 \, a_2 \, C\right ]$  goes to a constant, while for $a_2 \, C \, \to \, - \, \infty$, which we identify with the origin of the field manifold ($\mathfrak{z} \, \to \, 0$), the metric coefficient goes $\partial_C^2 J(C)$ to zero as $|\mathfrak{z}|^2 \, = \,  \exp \,\left [ 2 \, a_2 \, C\right ]$. This asymptotic behavior is essential for the interpretation of the shift $B\to B+ c$ as a compact rotation, as we pointed out before.
\paragraph{The flat $\mathbb{R}$ model.} In this case the metric coefficient  $\partial_C^2 J(C) \, = \, a_1^2 $  is constant everywhere and for $ C \, \to \, - \, \infty$ it does not go to zero. Such behavior selects the interpretation of the shift $B\to B+ c$ as a non-compact translation symmetry.

\subsection{Constant negative curvature models}
In eq. (3.16) of \cite{piesashatwo}  the general solution of the constant curvature equation:
\begin{equation}\label{formidabile}
    R(\phi) \, = \, - \, 4 \, \nu^2
\end{equation}
was presented in terms of the momentum map $\mathcal{P}(\phi)$ and of the canonical variable $\phi$. We have \footnote{Note that for the sake of our following arguments the solution of \cite{piesashatwo} is rewritten here in terms of exponentials rather than in terms of hyperbolic functions  $\cosh$ and $\sinh$.}:
\begin{equation}\label{gordino}
    \mathcal{P}(\phi) \, = \, a \, \exp(\nu \,\phi) \, + \,  b \, \exp (-\, \nu \,\phi) \, + \, c \quad ; \quad a,b,c \, \in \, \mathbb{R}
\end{equation}
In order to convert this solution in terms of the Jordan function $J(C)$ of the VP coordinate $C$, it is convenient to remark that, up to constant shift redefinitions and sign flips of the canonical variable $\phi \to \pm \phi + \kappa$, which leave its kinetic term invariant,  there are only three relevant cases:
\begin{description}
  \item[A)] $a \ne 0, \,  b \ne 0$ and $a/b >0$. In this case, up to an overall constant, we can just set:
  \begin{equation}\label{corrupziaA}
    \mathcal{P}(\phi) \, = \, \frac{1}{\nu} \,\left(\cosh (\nu \,\phi) \, + \, \mu \right) \quad \Rightarrow \quad V(\phi) \, \propto \,  \left(\cosh (\nu \,\phi) \, + \, \mu\right)^2
  \end{equation}
  \item[B)] $a \ne 0, \, b \ne 0$ and $a/b <0$. In this case we can just set:
  \begin{equation}\label{corrupziaB}
    \mathcal{P}(\phi) \, = \, \frac{1}{\nu}\, \left(\sinh (\nu \,\phi) \, + \, \mu\right )\quad \Rightarrow \quad V(\phi) \, \propto \,  \left(\sinh (\nu \,\phi) \, + \, \mu\right)^2
  \end{equation}
  \item[C)] $a \ne 0, \, b = 0$. In this case we can just set:
  \begin{equation}\label{corrupziaC}
    \mathcal{P}(\phi) \, = \, \frac{1}{\nu}\, \left(\exp (\nu \,\phi) \, + \, \mu \right) \quad \Rightarrow \quad V(\phi) \, \propto \,  \left(\exp (\nu \,\phi) \, + \, \mu\right)^2
  \end{equation}
\end{description}
\subsubsection{Elaboration of case A)}
\begin{figure}[!hbt]
\begin{center}
\iffigs
\includegraphics[height=70mm]{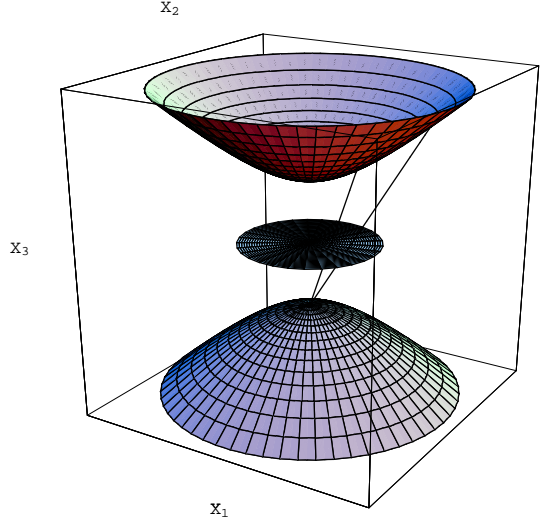}
\else
\end{center}
 \fi
\caption{\it  In this figure we show the hyperboloid ruled by lines of constant $\phi$ that are circles and of constant $B$ that are hyperbolae. In this figure we also show the stereographic projection of points of the hyperboloid onto points of the unit disk}
\label{hyperboloide}
 \iffigs
 \hskip 1cm \unitlength=1.1mm
 \end{center}
  \fi
\end{figure}

Let us  consider the case of the momentum map of eq.(\ref{corrupziaA}). The corresponding two-dimensional metric is:
\begin{equation}\label{giroscopioA}
    ds^2_\phi \, = \, \ft 14 \left(d\phi^2 \, + \, \sinh^2\left(\nu \,\phi\right) \, dB^2\right)
\end{equation}
which can be shown to be  the pull-back of the $(2,1)$-Lorentz metric onto a hyperboloid surface. Indeed setting:
\begin{eqnarray}
  X_1 &=& \sinh (\nu  \phi ) \cos (B \nu )\nonumber\\
  X_2 &=& \sinh (\nu  \phi ) \sin (B \nu ) \nonumber\\
  X_3 &=& \pm   \cosh (\nu  \phi ) \label{figliuttoA}
\end{eqnarray}
we obtain a parametric covering of the algebraic locus:
\begin{equation}\label{sistuloA}
    X_1^2 \, + \, X_2^2 \, - \, X_3^2 \, = \, - \, 1
\end{equation}
and we can verify that:
\begin{equation}\label{galloA}
    \frac{1}{4\nu^2} \, \left(dX_1^2 \, + \, dX_2^2 \, - \, dX_3^2 \right) \, = \, \ft 14 \left(d\phi^2 \, + \, \sinh^2\left(\nu \,\phi\right) \, dB^2\right) \, = \, ds^2_\phi
\end{equation}
A picture of the hyperboloid ruled by lines of constant $\phi$ and constant $B$ according to the parametrization (\ref{figliuttoA}) is depicted in fig.\ref{hyperboloide}.
Applying to the present case the general rule given in eq.(\ref{sodoma})  that defines the VP coordinate $C$ we get:
\begin{equation}\label{CifunzionoA}
    C(\phi) \, = \, \int \frac{d\phi}{\mathcal{P}^\prime(\phi)} \, = \,  \frac{\log \left(\tanh
   \left(\frac{\nu  \phi
   }{2}\right)\right)}{\nu ^2} \quad \Leftrightarrow \quad \phi = \frac{2 \mbox{Arctanh}
   \,\left(e^{C \nu
   ^2}\right)}{\nu }
\end{equation}
from which we deduce that the allowed range of the flat variable $C$, in which the canonical variable $\phi$ is real and goes from $0$ to $\infty$, is the following one:
\begin{equation}\label{dorolatteA}
    C \, \in \, \left[ -\, \infty \, , \, 0 \right ]
\end{equation}
Next applying  to the present case the general formula given in eq. (\ref{gartoccio}) that yields the  K\"ahler function $J(\phi)$ we obtain :
\begin{equation}\label{calleroA}
   J(\phi) \, = \, \int \frac{\mathcal{P}(\phi)}{\mathcal{P}^\prime(\phi)} \, d\phi \, = \, \frac{\log (\sinh (\nu  \phi
   ))}{\nu ^2}+\frac{\mu  \log
   \left(\tanh \left(\frac{\nu
   \phi }{2}\right)\right)}{\nu
   ^2}
\end{equation}
Substituting eq.(\ref{CifunzionoA}) into (\ref{calleroA}), after some manipulations we obtain:
\begin{equation}\label{goliardina}
    J(C) \, = \,  (\mu +1) \, C\,-\, \frac{\log
   \left(1-e^{2 C \nu
   ^2}\right)}{\nu ^2}+\frac{\log
   (2)}{\nu ^2}
\end{equation}
which corresponds to the following metric:
\begin{equation}\label{cucruto}
    ds^2_C \, = \, \ft 14 \, \frac{\partial^2 J(C)}{\partial C^2} \left(dC^2 \, + \, dB^2 \right) \, = \,\frac{1}{4}
   \left({dB}^2+{dC}^2\right) \nu ^2
   \mbox{csch}^2\left(C \nu^2\right)
\end{equation}
Upon use of the coordinate transformation (\ref{CifunzionoA}) the line element  $ds^2_C$ flows into $ds^2_\phi $ and viceversa.
\par
It remains to be seen how  such a metric is canonically written in terms of a complex coordinate $\mathfrak{z}\, = \,\zeta$. In this case the appropriate relation between $\zeta$ in the unit circle and the real variables $C,B$ is the following:
\begin{equation}\label{zetosaA}
   \zeta \, = \,  e^{\nu^2  ({\rm i} B  +C )}
\end{equation}
With this position we find:
\begin{equation}\label{coccolino}
  ds^2_C \, = \,  \frac{1}{\nu^2} \, \frac{d\zeta \, d\bar{\zeta}}{\left( 1\, - \,\zeta \, \bar{\zeta} \right)^2} \, = \, \,\partial_\zeta\, \partial_{\bar{\zeta}}\, \underbrace{ \left[-\,\frac{1}{\nu^2} \,\log\left( 1\, - \,\zeta \, \bar{\zeta}\right)+ \frac{\mu + 1}{2 \, \nu^2} \log|\zeta \, \bar{\zeta}|^2\right]}_{J(C) \, = \,\mathcal{K}(\zeta\, , \, \bar{\zeta})\, = \,\mbox{K\"ahler potential}} \,d\zeta \, d\bar{\zeta}
\end{equation}
On the other hand from the position (\ref{zetosaA}) it is evident that the shift in $B$ is a compact rotation of the complex coordinate $\zeta$.
\par
For $C \to - \infty$, namely for  large and negative values of the argument we have the following expansion of the K\"ahler potential
\begin{eqnarray}
  J(C) & = &  \frac{\log(2)}{\nu ^2}\, + \,(\mu +1) \, C \, + \, \frac{e^{2 C \nu^2}}{\nu ^2}+\frac{e^{4 C \nu^2}}{2 \nu ^2}+\frac{e^{6 C\nu ^2}}{3 \nu ^2}
    + \mathcal{O}(e^{8 C\nu ^2})\nonumber\\
   \null &\Downarrow & \null \nonumber\\
   \partial_C^2 J(C) & = &  4 \, {\nu ^2} \,  e^{2 C \nu^2}\, + \, 8 \,\nu ^2 \,e^{4 C \nu^2}\, + \, 12 \,\nu ^2 \,e^{6 C \nu^2}\,+ \mathcal{O}(e^{8 C\nu ^2}) \, \stackrel{C\,\to \, -\, \infty}{\Longrightarrow}\, 0\label{minchiusA}
\end{eqnarray}
while for $C \to 0$, which corresponds to the boundary of moduli space, we have a logarithmic singularity:
\begin{equation}\label{ciurlacca}
 J(C) \, = \,  \frac{\log(2)}{\nu ^2} \,- \,\frac{\log \left(-2 \nu^2\right)}{\nu ^2}\, -\frac{\log (C)}{\nu^2} \, + \, \mu \, C \,  -\frac{1}{6} C^2 \nu ^2 \, + \, \mathcal{O}(C^3)
\end{equation}
The interpretation of the parameter $\mu$ is evident from the above formulae. It introduces a Fayet-Iliopoulos term.
\par
It is useful to write the $D$-type scalar potential in three different forms, as function of the canonical field $\phi$, as function of the VP coordinated $C$ and as function of the complex coordinate $\zeta$:
\begin{equation}\label{sicumera}
    V \, \propto \, \left( \cosh\left(\nu \, \phi \right) \, + \, \mu\right)^2 \, = \, \left(\mu +\frac{2 e^{2 C
   \nu ^2}}{1-e^{2 C \nu
   ^2}}+1\right)^2\, = \,\frac{1}{\nu^4} \, \left( \frac{\mu+1 \, - \, \mu \, \zeta \, \bar{\zeta}}{1 \, - \, \zeta \, \bar{\zeta}}\right)^2
\end{equation}
\begin{figure}[!hbt]
\begin{center}
\iffigs
\includegraphics[height=50mm]{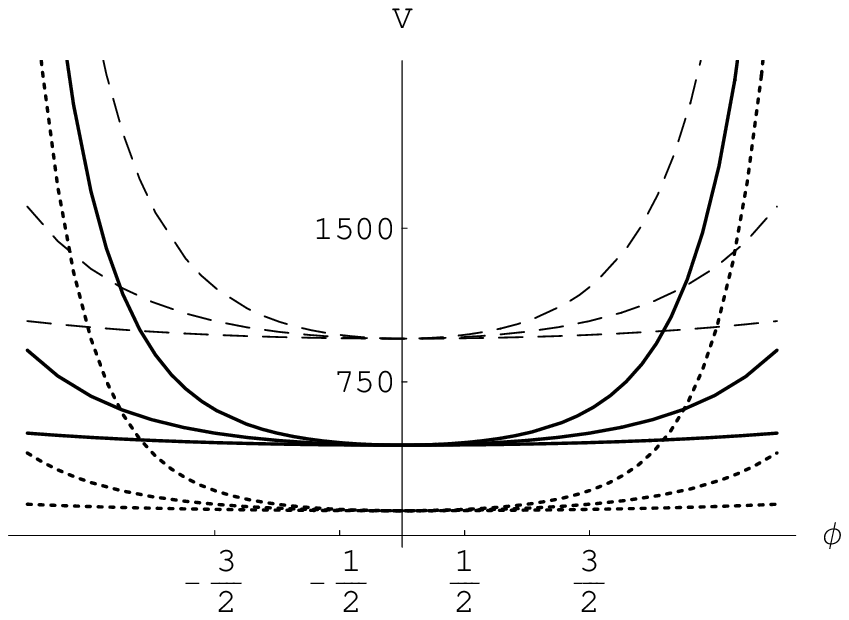}
\includegraphics[height=50mm]{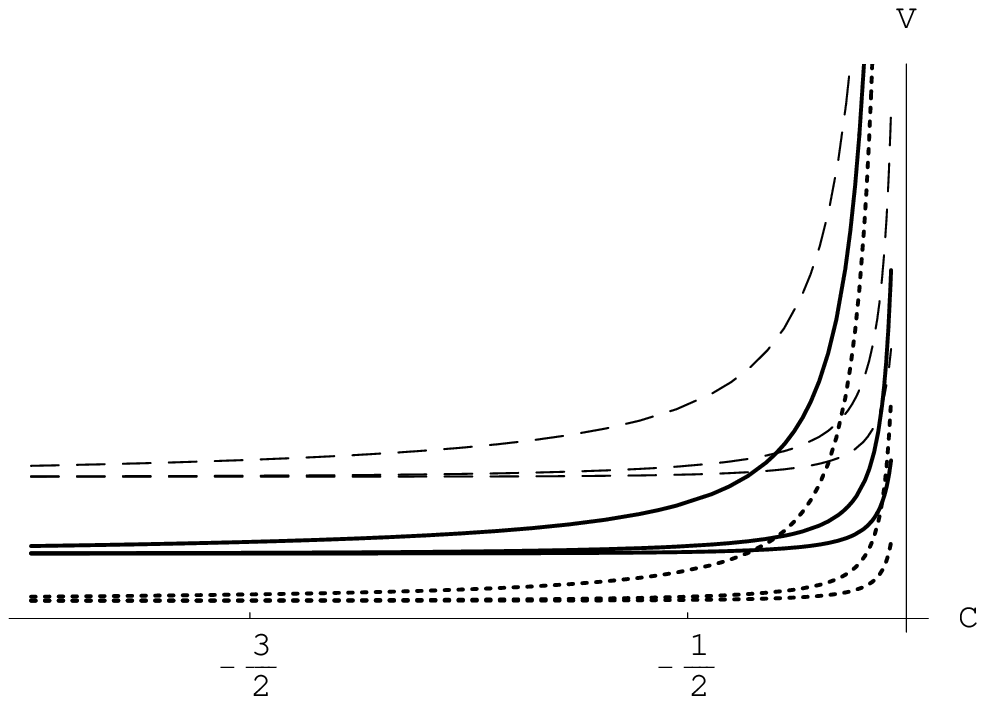}
\includegraphics[height=50mm]{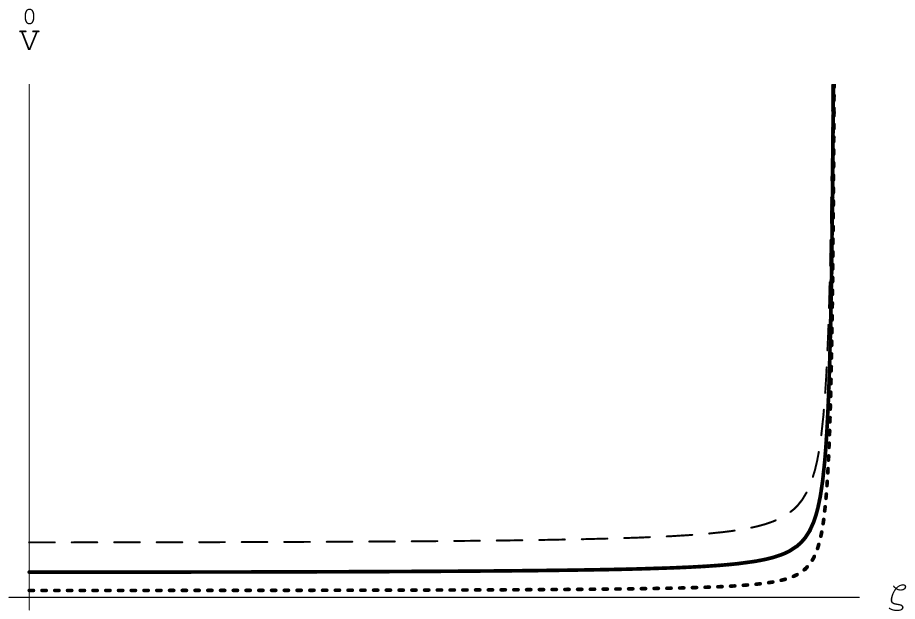}
\else
\end{center}
 \fi
\caption{\it Plots of the  inflaton potential corresponding to the gauging of a compact $U(1)$ isometry inside $\mathrm{SL(2,\mathbb{R})}$. The potential is displayed as function of  either the canonical coordinate $\phi$ (first picture) or the VP coordinate $C$ (second picture) or the modulus of the complex coordinate $\zeta$ according to the formulae displayed in eq.(\ref{sicumera}). In each picture the curves of the same type (solid, dashed or long dashed) correspond to the same value of the Fayet Iliopoulos parameter $\mu$, but to different values of the curvature parameter $\nu$. }
\label{potenzialone}
 \iffigs
 \hskip 1cm \unitlength=1.1mm
 \end{center}
  \fi
\end{figure}

\subsubsection{Elaboration of case B)}
\begin{figure}[!hbt]
\begin{center}
\iffigs
\includegraphics[height=70mm]{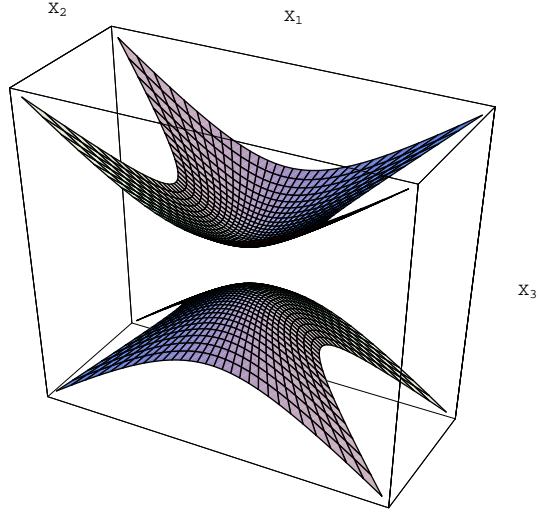}
\caption{\it  The hyperboloid surface displayed in the parametrization (\ref{figliutto}). The lines drawn on the hyperboloid surface are those of constant $B$ and constant $\phi$ respectively. Both of them are  hyperbolae, in this case. }
\label{iperbolonetwo}
 \iffigs
 \hskip 1cm \unitlength=1.1mm
 \end{center}
  \fi
\end{figure}

Let us now consider the case of the momentum map of eq.(\ref{corrupziaB}). The corresponding two-dimensional metric is:
\begin{equation}\label{giroscopio}
    ds^2_\phi \, = \, \ft 14 \left(d\phi^2 \, + \, \cosh^2\left(\nu \,\phi\right) \, dB^2\right)
\end{equation}
which can be shown to be another form of the pull-back of the Lorentz metric onto a hyperboloid surface. Indeed setting:
\begin{eqnarray}
  X_1 &=& \cosh (\nu  \phi ) \sinh (B \nu )\nonumber\\
  X_2 &=& \sinh (\nu  \phi ) \nonumber\\
  X_3 &=& \pm \cosh (B \nu ) \cosh (\nu  \phi ) \label{figliutto}
\end{eqnarray}
we obtain a parametric covering of the algebraic locus:
\begin{equation}\label{sistulo}
    X_1^2 \, + \, X_2^2 \, - \, X_3^2 \, = \, - \, 1
\end{equation}
and we can verify that:
\begin{equation}\label{gallo}
    \frac{1}{4\nu^2} \, \left(dX_1^2 \, + \, dX_2^2 \, - \, dX_3^2 \right) \, = \, \ft 14 \left(d\phi^2 \, + \, \cosh^2\left(\nu \,\phi\right) \, dB^2\right) \, = \, ds^2_\phi
\end{equation}
A three-dimensional picture of the hyperboloid ruled by lines of constant $\phi$ and constant $B$ is displayed in fig.\ref{iperbolonetwo}.

Applying to the present case the general rule given in eq. (\ref{sodoma})  that defines the VP coordinate $C$ we get:
\begin{equation}\label{Cifunziono}
    C(\phi) \, = \, \int \frac{d\phi}{\mathcal{P}^\prime(\phi)} \, = \, \frac{2 \mbox{Arctan} \left(\tanh
   \left(\frac{\nu  \phi
   }{2}\right)\right)}{\nu ^2} \quad \Leftrightarrow \quad \phi = \frac{2 \mbox{Arctanh}
   \left(\tan \left(\frac{C
   \nu ^2}{2}\right)\right)}{\nu}
\end{equation}
from which we deduce that the allowed range of the flat variable $C$, in which the canonical variable $\phi$ is real and goes from $-\infty$ to $\infty$, is the following one:
\begin{equation}\label{dorolatte}
    C \, \in \, \left[ -\, \frac{\pi}{2 \, \nu^2} \, , \, \frac{\pi}{2 \, \nu^2} \right ]
\end{equation}
Next applying  to the present case the general formula given in eq. (\ref{gartoccio})  that yields the  K\"ahler function $J(\phi)$ we obtain :
\begin{equation}\label{calleroB}
   J(\phi) \, = \, \int \frac{P(\phi)}{\mathcal{P}^\prime(\phi)} \, d\phi \, = \, \frac{2 \mu  \mbox{Arctan}
   \,\left(\tanh
   \left(\frac{\nu  \phi
   }{2}\right)\right)}{\nu
   ^2}+\frac{\log (\cosh (\nu
   \phi ))}{\nu ^2}
\end{equation}
Substituting eq.(\ref{Cifunziono}) into (\ref{calleroB}), after some manipulations we obtain:
\begin{equation}\label{goliardina}
    J(C) \, = \,  \mu \, C \,- \,\frac{1}{\nu ^2} \,\log
   \left(\cos \left(C \nu
   ^2\right)\right)
\end{equation}
which corresponds to the following metric:
\begin{equation}\label{cucruto}
    ds^2_C \, = \, \ft 14 \, \frac{\partial^2 J(C)}{\partial C^2} \left(dC^2 \, + \, dB^2 \right) \, = \,\frac{1}{4}
   \left(\mbox{d}B^2+\mbox{d}C^2\right) \nu ^2 \sec ^2\left(C\nu ^2\right)
\end{equation}
Upon use of the coordinate transformation (\ref{Cifunziono}) the line element  $ds^2_C$ flows into $ds^2_\phi $ and viceversa.
\par
It remains to be seen how  such a metric is canonically written in terms of a complex coordinate $\zeta$. In this case the appropriate relation between $\zeta$ in the unit circle and the real variables $C,B$ is different. Setting:
\begin{equation}\label{zetosa}
   \zeta \, = \, {\rm i} \tanh \left(\frac{1}{2} (B-{\rm i}\, C) \nu ^2\right)
\end{equation}
which implies:
\begin{eqnarray}\label{golingus}
    C & = & -\, \frac{{\rm i}}{\nu^2} \, \left( \mbox{ArcTanh}(-\,{\rm i}\, \zeta)\, - \, \mbox{ArcTanh}(\,{\rm i}\, \bar{\zeta})\right)\nonumber\\
    B & = &  \frac{1}{\nu^2} \, \left( \mbox{ArcTanh}(-\,{\rm i}\, \zeta)\, + \, \mbox{ArcTanh}(\,{\rm i}\, \bar{\zeta})\right)
    \label{golingus}
\end{eqnarray}
we find:
\begin{equation}\label{durlacco}
   J(C)\, = \,  \mathcal{K}(\zeta,\bar{\zeta})\, = \,  \, - \, {\rm i} \, \frac{\mu}{\nu^2} \, \left( \mbox{ArcTanh}(-\,{\rm i}\, \zeta)\, - \, \mbox{ArcTanh}(\,{\rm i}\bar{\zeta})\right)\, - \, \frac{1}{\nu^2} \, \log\left(1 \, - \, \zeta \, \bar{\zeta}\right)
\end{equation}

\begin{equation}\label{coccolino}
  ds^2_C \, = \,  \frac{1}{\nu^2} \, \frac{d\zeta \, d\bar{\zeta}}{\left( 1\, - \,\zeta \, \bar{\zeta} \right)^2} \, = \, \,\partial_\zeta\, \partial_{\bar{\zeta}}\, \mathcal{K}(\zeta,\bar{\zeta}) \,d\zeta \, d\bar{\zeta}
\end{equation}
The identification (\ref{coccolino}) allows us to understand the nature of the isometry $B \, \to \, B +c$ which is non compact. To this effect let us  consider the image of the $\mathrm{SL(2,\mathbb{R})}$ dilatations inside the $\mathrm{SU(1,1)}$ group:
\begin{equation}\label{pittorello}
    \Lambda_\rho \, = \, \left(
\begin{array}{ll}
 1 & 1 \\
 {\rm i} & -{\rm i}
\end{array}
\right) \, \left(
\begin{array}{ll}
 e^{\rho } & 0 \\
 0 & e^{-\rho }
\end{array}
\right) \, \left(
\begin{array}{ll}
 \frac{1}{2} & -\frac{\rm i}{2} \\
 \frac{1}{2} & \frac{\rm i}{2}
\end{array}
\right) \, = \, \left(
\begin{array}{ll}
 \cosh \rho  & -\rm i \sinh \rho
   \\
 {\rm i} \sinh \rho  & \cosh \rho
\end{array}
\right)
\end{equation}
The action of this group on the complex coordinate $\zeta$ inside the unit circle is given by the following linear fractional transformation:
\begin{equation}\label{lineatafratta}
     \Lambda_\rho \cdot \zeta \, = \, \frac{\zeta  \cosh (\rho )\,-\,{\rm i}\, \sinh
   (\rho )}{\cosh (\rho )\,+ \,{\rm i}\, \zeta
   \sinh (\rho )}
\end{equation}
For an infinitesimal parameter $\rho \ll 1$ we have:
\begin{equation}\label{sinio}
    \zeta \, \to \, \zeta \, + \, \delta_\rho\zeta \quad ; \quad \delta_\rho\zeta \, = \, -{\rm i }\,\left(1 \, + \, \zeta^2\right) \rho
\end{equation}
Consider next the effect of a shift of $B$ on the complex coordinate $\zeta$ as given in eq.(\ref{zetosa}). We find:
\begin{equation}\label{giuseppe}
  \partial_B \, \zeta \, = \,   \frac{1}{2}\, \nu ^2 \, \,{\rm i} \,
   \mbox{sech}^2\left(\frac{1}{2}
   (B-{\rm i} \,C) \nu ^2\right) \, = \, \frac{1}{2} \, \nu ^2 \, {\rm i} \, \left(1 \, + \, \zeta^2\right)
\end{equation}
This shows that indeed the $B$-shifts realize the action of the non compact subgroup (\ref{lineatafratta}) on the complex coordinate $\zeta$.
\par
\begin{figure}[!hbt]
\begin{center}
\iffigs
\includegraphics[height=50mm]{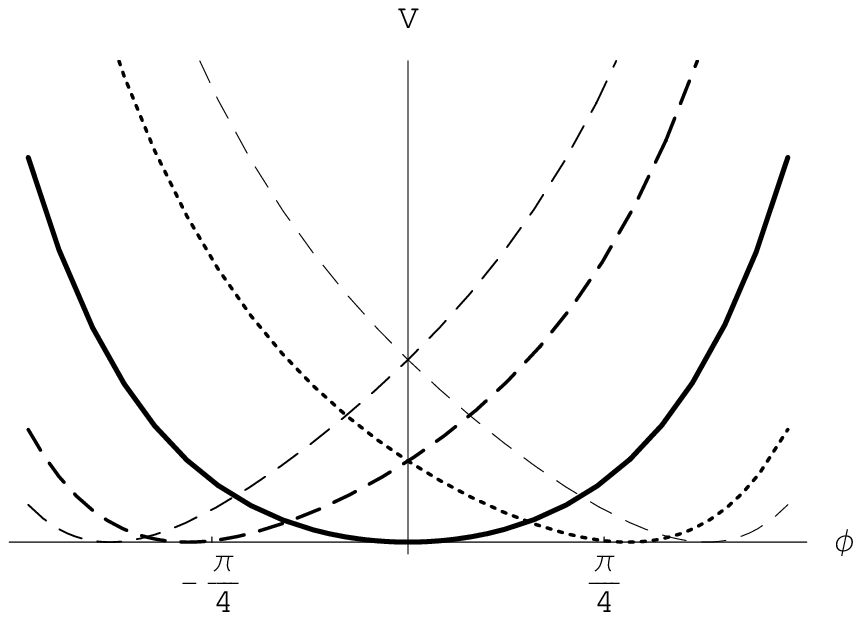}
\includegraphics[height=50mm]{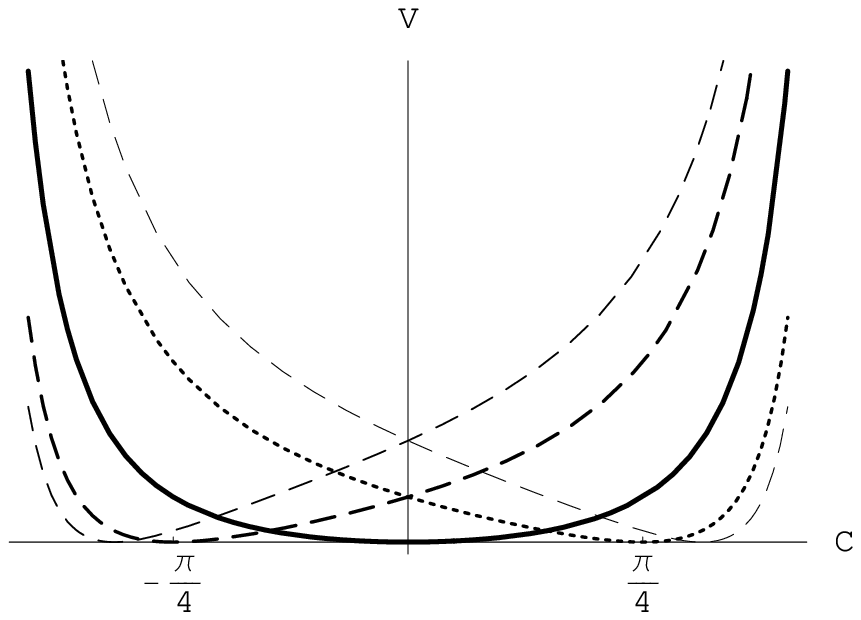}
\else
\end{center}
 \fi
\caption{\it Plots of the  inflaton potential corresponding to the gauging of a non-compact $\mathrm{SO(1,1)}$ isometry inside $\mathrm{SL(2,\mathbb{R})}$. The potential is displayed as function of  either the canonical coordinate $\phi$ (first picture) or the VP coordinate $C$ (second picture). In each picture the curves of different  type (solid, dashed or long dashed) correspond to different values of the Fayet Iliopoulos parameter $\mu$, and to the same value $\nu$ of the curvature parameter. }
\label{tangentalo}
 \iffigs
 \hskip 1cm \unitlength=1.1mm
 \end{center}
  \fi
\end{figure}

Knowing the Killing vector we can now write the scalar potential in three different ways as function of the canonical coordinate $\phi$, of the VP coordinate $C$ and of the complex coordinate $\zeta$. We find:
\begin{equation}\label{potinis}
 V \, \propto \, \left(\sinh(\nu \phi) +\mu\right)^2 \, = \, \left(\mu +\tan \left(C \nu
   ^2\right) \right)^2 \, =\,   \,\left(\frac{\mbox{$\bar \zeta $} (\zeta
   +\mbox{$\bar \zeta $}) \zeta
   +\zeta +\mbox{$\bar \zeta $}+2
   \mu  (\zeta  \mbox{$\bar \zeta $}-1)}{4 \zeta
   \mbox{$\bar \zeta $}-4}\right)^2
\end{equation}
The behavior of this family of potentials is displayed in fig.\ref{tangentalo}.
Note that when written in terms of the complex variable $\zeta$ the potential does not appear to depend only on one variable. Yet this is so since the potential depends only from the $C$-variable defined by eq. (\ref{golingus}).
\paragraph{Asymptotic behavior}
It is now important to consider the behavior of the K\"ahler function $J(C)$ as given in eq.(\ref{goliardina}) when the VP coordinate approaches the boundary of its own range. The point $C=0$ is perfectly regular for $J(C)$ and indeed, in consideration of eq.(\ref{dorolatte}), it is well inside the range of definition. The boundary is approached when $C \, \to \, \pm \frac{\pi}{2 \,\nu^2 }$. For this reason we set $C=\frac{\pi}{2 \,\nu^2 }-\xi$ and we consider the behavior of the K\"ahler function for $\xi\simeq0$. We obtain:
\begin{equation}\label{tarlinetto}
  J(\xi) \, = \,   -\xi  \mu +\frac{\pi  \mu
   }{2 \nu ^2}-\frac{\log
   \left(\nu ^2\right)}{\nu
   ^2}-\frac{\log (\xi )}{\nu ^2} \, + \, \mathcal{O}(\xi^2)
\end{equation}
In this way we put into evidence the logarithmic singularity which characterizes the behavior of the K\"ahler function in at the boundary. Once again it also appears that the parameter $\mu$ plays the role of Fayet Iliopoulos term. Considering now the behavior of the function $\partial_C^2 J(C)$ for $C \, \to \, 0$, which is the center of the bulk for the field manifold, we find:
\begin{equation}\label{cruscotto}
  \partial_C^2 J(C) \, \stackrel{C \,\to \, 0}{\simeq} \, \nu ^2+\nu ^6 C^2+\frac{2 \nu
   ^{10}
   C^4}{3}+\mathcal{O}\left(C^5\right)
\end{equation}
We see that the metric coefficient goes to a constant and this is the obstacle to interpret the symmetry $B\to B +c$ as a compact rotation. Indeed as we have seen it is rather a hyperbolic transformation.
\subsubsection{Elaboration of case C)}
In the case the momentum map is given by eq.(\ref{corrupziaC}) by immediate integration of eq.(\ref{sodoma}) we  obtain the VP coordinate $C(\phi)$ and its inverse function:
\begin{equation}\label{siccatus}
    C(\phi) \, = \, -\frac{e^{-\nu  \phi }}{\nu ^2} \, \Leftrightarrow \, \phi(C) \, = \,-\frac{\log
   \left(-C \nu ^2\right)}{\nu
   }
\end{equation}
The integration of eq.(\ref{gartoccio}) for the K\"ahler potential is equally immediate and we find:
\begin{equation}\label{filantropone}
    J(\phi) \, = \,\frac{\phi -\frac{e^{-\nu  \phi
   } \mu }{\nu }}{\nu } \quad \Leftrightarrow \quad J(C) \, = \, \mu \, C \, -\, \frac{1}{\nu ^2} \,\log \left(-C
   \right)\, + \, \mbox{const}
\end{equation}
From the form of equation (\ref{filantropone}) we conclude that the appropriate solution of the complex structure equation in this case is:
\begin{equation}\label{curiaceo}
    \mathfrak{z} \, = \, t \, = \, \, {\rm i} C \, - \, B
\end{equation}
so that the K\"ahler metric becomes proportional to the Poincar\'e metric in the upper complex plane (note that $C$ is negative definite for the whole range of the canonical variable $\phi$):
\begin{equation}\label{cicculini}
    ds^2 \, = \, \ft 14 \, \frac{\mathrm{d}^2J}{\mathrm{d}C^2} \, \left(dC^2 + dB^2\right)\, = \, \frac{1}{4\, \nu^2} \, \frac{d\bar{t} \, dt}{\left(\mbox{Im}t\right)^2}
\end{equation}
As a consequence of equation (\ref{curiaceo}), we see that the $B$-translation happens to be, in this case, a non-compact shift symmetry.
\par
As in the previous cases we can write the potential in three forms:
\begin{equation}\label{calzina}
    V\, = \, \left( \exp[\nu \, \phi ] \, + \, \mu\right)^2 \, = \, \left(\mu \, + \, \frac{1}{\nu^2 \, C}\right)^2 \, = \, \left( \ft 12 \, \mu \, + \, \frac{{\rm i}}{\nu^2} \, \left(t-\bar{t}\right)^{-1}\right)^2
\end{equation}
The results for the five type of potentials that we have obtained from constant curvature symmetric spaces are summarized
in table \ref{potenziallini}.
\begin{figure}[!hbt]
\begin{center}
\iffigs
\includegraphics[height=50mm]{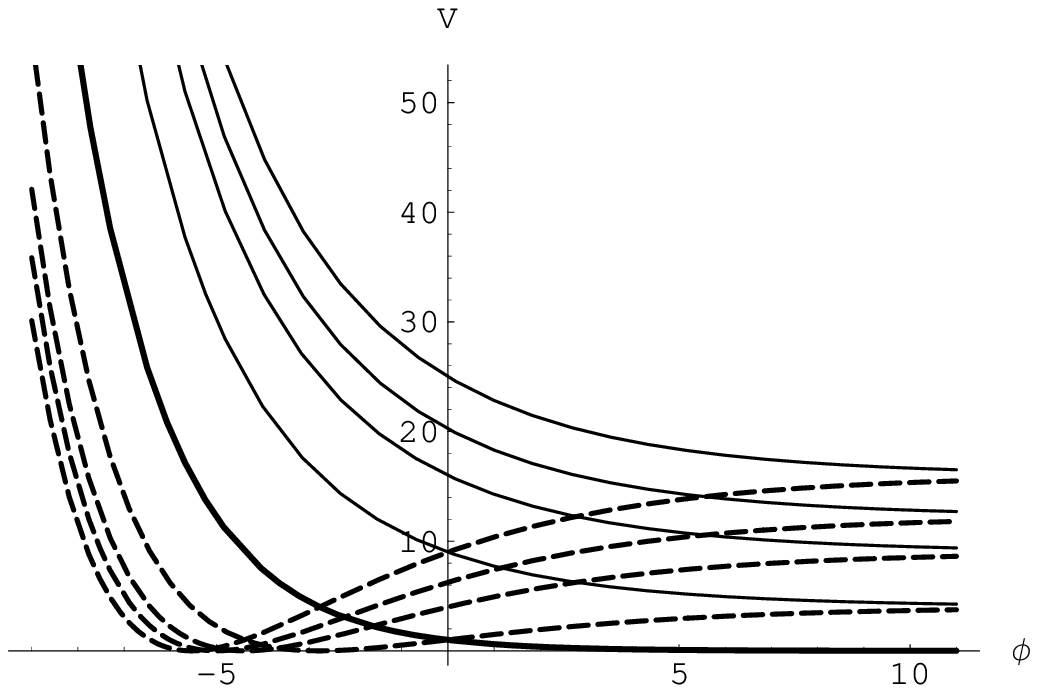}
\includegraphics[height=50mm]{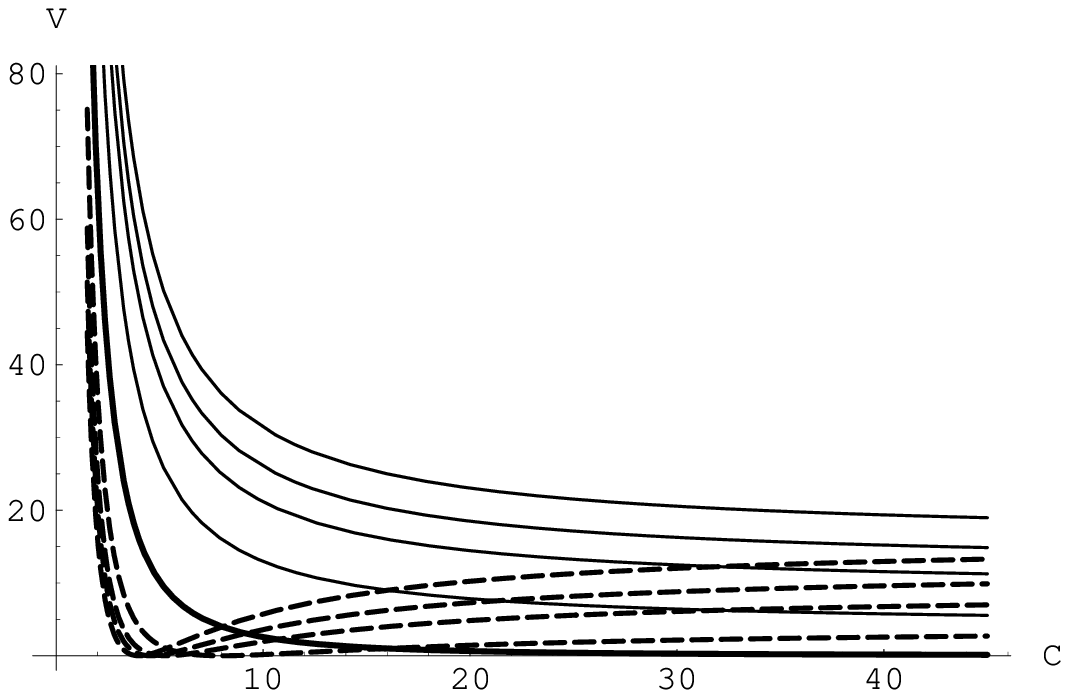}
\else
\end{center}
 \fi
\caption{\it Plots of the  inflaton potential corresponding to the gauging of a shift-isometry (parabolic) inside $\mathrm{SL(2,\mathbb{R})}$ (Starobinsky-like models). The potential is displayed as a function of  either the canonical coordinate $\phi$ (first picture) or of the VP coordinate $C$ (second picture). In each picture the solid lines correspond to positive values of the Fayet Iliopoulos parameter $\mu$, while the dashed lines correspond to negative values of $\mu$, which develop a minimum at $V=0$. The value of $\nu$ is the same in all cases and it is negative $\nu \, = \, \ft 14$. }
\label{starobino}
 \iffigs
 \hskip 1cm \unitlength=1.1mm
 \end{center}
  \fi
\end{figure}
\begin{table}[h!]
{\scriptsize
\centering
% centering table
\begin{tabular}{|c|c|c|c|c|c|}
\hline
Curv. & Gauge Group & $V(\phi)$ & $V(C)$ & $V(\mathfrak{z} )$ & Comp. Struct. \\
\hline
\hline
\null & \null & \null & \null & \null & \null \\
$-\nu^2$ & $\mathrm{U(1)}$ & $\left( \cosh\left(\nu \, \phi \right) \, + \, \mu\right)^2 $ & $ \left(\mu +\frac{2 e^{2 C
   \nu ^2}}{1-e^{2 C \nu
   ^2}}+1\right)^2$ & $\frac{1}{\nu^4} \, \left( \frac{\mu+1 \, - \, \mu \, \zeta \, \bar{\zeta}}{1 \, - \, \zeta \, \bar{\zeta}}\right)^2$ & $\zeta \, = \, e^{C \, - \, {\rm i}B}$  \\
\null & \null & \null & \null & \null & \null \\
\hline
\null & \null & \null & \null & \null & \null \\
$-\nu^2$ & $\mathrm{SO(1,1)}$ & $\left( \sinh\left(\nu \, \phi \right) \, + \, \mu\right)^2 $ & $ \left(\mu +\tan \left(C \nu
   ^2\right) \right)^2 $&   $ \left(\frac{{\bar \zeta} (\zeta
   + {\bar \zeta }) \zeta
   +\zeta +{\bar \zeta}+2
   \mu  (\zeta  \,{\bar \zeta}-1)}{4 \zeta \,{\bar \zeta}\,-\,4}\right)^2 $ &
   $\zeta \, = \, {\rm i}\, \tanh \left(\ft 12 (B-{\rm i}C)\, \nu^2 \right)$ \\
\null & \null & \null & \null & \null & \null \\
\hline
\null & \null & \null & \null & \null & \null \\
$-\nu^2$ & $\mathrm{parabolic}$ & $\left( \exp\left(\nu \, \phi \right) \, + \, \mu\right)^2 $ & $ \left(\mu \, + \, \frac{1}{\nu^2 \, C}\right)^2 $ & $ \left( \ft 12 \, \mu \, + \, \frac{{\rm i}}{\nu^2} \, \left(t-\bar{t}\right)^{-1}\right)^2 $ &
   $t \, = \, - {\rm i}\,C \, + \, B$ \\
\null & \null & \null & \null & \null & \null \\
\hline
\null & \null & \null & \null & \null & \null \\
$0$ & $\mathrm{U(1)}$ & $M^4 \left[ \left(\frac{\phi}{\phi_0} \right)^2 \, \pm \, 1 \right]^2$ &  $M^4 \left[
\frac{e^{2 a_2 C}}{\phi_0^2}  \, \pm \, 1 \right]^2$  & $ \frac{1}{4} \, \left(  \mathfrak{z} \, \bar{\mathfrak{z}} \, - \, \frac{2\, a_0}{a_2} \, \right)^2 $ &
   $\mathfrak{z} \, = \, \exp \left [ a_2 \left(C \, + \, {\rm i} B\right)\right]$  \\
\null & \null & \null & \null & \null & \null \\
\hline
\null & \null & \null & \null & \null & \null \\
$0$ & $\mathrm{parabolic}$ & $\left(a_0 \, + \, a_1\, \phi\right)^2$ &  $\left(a_1 \, C \, +\, \beta\right)^2$  & $ \frac{1}{2} \, \left(  a_1 \mbox{Im} \mathfrak{z}  \, + \beta \right)^2 $ &
   $\mathfrak{z} \, = \, {\rm i} C \, + \, B$  \\
\null & \null & \null & \null & \null & \null \\
\hline
\end{tabular}
}
\caption{Summary of the potentials of $D$-type  obtained from constant curvature K\"ahler manifolds by gauging either a compact or a non compact isometries}
\label{potenziallini}
\end{table}

\section{Conclusions}
Summarizing, the main result of the present paper concerns three related points:
\begin{description}
  \item[A)] The physical properties of the minimal supergravity models that encode one-field cosmologies with a positive definite potential depend in a crucial way on the global topology of the group $\mathcal{G}$ that is gauged in order to produce them. When it is compact we have a certain pattern of symmetries and charge assignments, when it is non-compact we have a different pattern.
  \item[B)] The global topology of the group $\mathcal{G}$ reflects into a different asymptotic behavior of the function $\partial_C^2 J(C)$ in the region that we can call the origin of the manifold. In the compact case the complex field $\mathfrak{z}$ is charged with respect to $\mathrm{U(1)}$ and, for consistency, this symmetry should exist at all orders in an expansion of the scalar field $\sigma$-model for small fields. Hence for $\mathfrak{z} \to 0$ the kinetic term of the scalars should go to the standard canonical one:
      \begin{equation}\label{gomoroid}
      \mathcal{L}^{(can)}_{kin} \, = \, \ft 14  \partial_\mu \mathfrak{z} \, \partial^\mu \bar{ \mathfrak{z}}
      \end{equation}
      Assuming, as it is necessary for the $\mathrm{U(1)}$ interpretation of the $B$-shift symmetry, that $\mathfrak{z} \, = \, \zeta \, \, = \, \exp\left[\alpha({\rm i} C \, - \, B)\right]$, where $\alpha$ is some real coefficient, eq.(\ref{gomoroid}) can be satisfied if and only if we have:
      \begin{equation}\label{canolicchio}
        \lim_{C \, \to \, - \, \infty} \, \exp\left[ - \, 2 \, \alpha \, C\right] \, \partial_C^2 J(C) \, = \, \mbox{const}
      \end{equation}
      which therefore is an intrinsic clue to establish the global topology of the inflaton K\"ahler surface $\Sigma$.
      \item[C)] The Fayet Iliopoulos terms always identified as linear terms in VP coordinate $C$ in the function $J(C)$ are rather different in the complex variable $\mathfrak{z}$, depending on which is the appropriate topology.
\end{description}
These properties are general and apply to all inflaton models embedded into   a minimal $\mathcal{N}=1$ supergravity description. In the particular case of constant cruvature K\"ahler surfaces we were able to derive five models, two associated with a flat K\"ahler manifold and three with the unique negative curvature two-dimensional symmetric space $\mathrm{SL(2,R)/O(2)}$. Of these five models three correspond to known inflationary potentials: the Higgs potential and  the chaotic inflation  quadratic potential, coming from a zero curvature K\"ahler manifold  and the Starobinsky-like potentials, coming  from the gauging of parabolic subgroups of $\mathrm{SL(2,R)}$. These latter potentials were already embedded in supergravity in \cite{minimalsergioKLP} and \cite{Farakos:2013cqa}. The remaining two potentials, respectively associated with the gauging of elliptic and hyperbolic subgroups so far have not yet been utilized as candidate inflationary potentials and possible they are incompatible with PLANCK data. In any case it is important to emphasize that parabolic Starobinsky-like potentials are associated with higher curvature supergravity models (\cite{Cecotti:1987qe},\cite{Cecotti:1987sa},\cite{contownsend},\cite{D'Auria:1988qm}) and it is an obvious question to inquiry what is the origin, in this context, of the elliptic and hyperboic Starobinsky-like potentials we have found.
Furthermore let us stress that the Fayet Iliopoulos term (and its sign)
drastically changes the behavior of the scalar potential. In the
Starobinsky case it is responsable for the de Sitter inflationary  phase.
It is furthermore interesting to note that for some particular values of the curvature and of the Fayet Iliopoulos parameter the models classified in this paper become integrable.
In  table \ref{integpotenziallini} we list such cases. They are in the intersection of the list of table \ref{potenziallini} with  the list of integrable series of potentials classified in \cite{noicosmoitegr} and further analyzed in \cite{piesashatwo}.
\begin{table}[h!]
\begin{center}
{\scriptsize
% centering table
\begin{tabular}{|c|c|c|c|c|c|}
\hline
Curv. & Gauge Group & $V(\phi)$ & Values of $\nu$  & Values of $\mu$ & Mother series \\
\hline
\hline
\null & \null & \null & \null & \null & \null \\
$-\nu^2$ & $\mathrm{U(1)}$ & $\left( \cosh\left(\nu \, \phi \right) \, + \, \mu\right)^2 $ & $\nu \, = \, \frac{\sqrt{3}}{2}$ & $\mu \, = \,0$ & $I_1$ or $I_7$ with $\gamma \, = \, \ft 12$  \\
\null & \null & \null & \null & \null & \null \\
\hline
\null & \null & \null & \null & \null & \null \\
$-\nu^2$ & $\mathrm{U(1)}$ & $\left( \cosh\left(\nu \, \phi \right) \, + \, \mu\right)^2 $ & $\nu \, = \, \frac{2}{\sqrt{3}}$ & $\mu \, = \,1$ &   $I_7$ with $\gamma \, = \, \ft 13$  \\
\null & \null & \null & \null & \null & \null \\
\hline
\null & \null & \null & \null & \null & \null \\
$-\nu^2$ & $\mathrm{U(1)}$ & $\left( \cosh\left(\nu \, \phi \right) \, + \, \mu\right)^2 $ & $\nu \, = \, \frac{2}{\sqrt{3}}$ & $\mu \, = \, - \,1$ &   $I_7$ with $\gamma \, = \, \ft 13$  \\
\null & \null & \null & \null & \null & \null \\
\hline
\null & \null & \null & \null & \null & \null \\
$-\nu^2$ & $\mathrm{SO(1,1)}$ & $\left( \sinh\left(\nu \, \phi \right) \, + \, \mu\right)^2 $ & $\nu \, = \, \frac{\sqrt{3}}{2}$ & $\mu \, = \,0$ & $I_1$ or $I_7$ with $\gamma \, = \, \ft 12$  \\
\null & \null & \null & \null & \null & \null \\
\hline
\null & \null & \null & \null & \null & \null \\
$-\nu^2$ & $\mathrm{parabolic}$ & $\left( \exp\left(\nu \, \phi \right) \, + \, \mu\right)^2 $ & $\nu \, = \, \mbox{any}$ & $\mu \, = \,0$ & all pure exp are integ.  \\
\null & \null & \null & \null & \null & \null \\
\hline
\end{tabular}
}
\end{center}
\caption{In this table we mention which particular values of the curvature and of the Fayet Iliopoulos constant yield cosmological potentials that are both associated to constant curvature and integrable according to the classification of \cite{noicosmoitegr}}
\label{integpotenziallini}
\end{table}
By means of  the arguments contained in this paper we have emphasized the physical relevance of the global topology of the $\Sigma$ K\"ahler surface associated with minimal supergravity models of inflations. Global topology amount at the end of the day to giving the precise range of the coordinates $C$ and $B$ labeling the points of $\Sigma$. In the five constant curvature cases we presented these ranges are as follows.
 In the elliptic and parabolic case  $C$  is in the range $[-\infty,-0]$ in the elliptic and
while it is in the range  $[-\infty,+\infty]$  for the flat case and it is periodic in the hyperbolic case.
The cooordinate $B$ instead is periodic in the elliptic case, it is unrestricted in the hyperbolic and parabolic cases. The flat case with $B$ periodic is a strip. It is instead the full plane in the parabolic coordinate.
\par
Finally let us stress that the considerations put forward here can be extended to a large class of inflationary models based on non symmetric spaces, namely associated with K\"ahler surfaces $\Sigma$ whose curvature is non-constant. Among them a subclass of models are the integrable ones for which a preliminary analysis was given in \cite{piesashatwo}. In a forthcoming publication \cite{pietrosashasergioTwo} we plan to extend and improve such analysis for many models, also of realistic type both non integrable and occasionally integrable.
\section*{Acknowledgments}
%%%%%%%%%%%%%%%%%%%%%%%%%%%%%%
One of us (S.F.) would like to aknowledge enlightening discussions with R. Kallosh, A. Linde and M. Porrati on related work.
\par
S.F. is supported by ERC Advanced Investigator Grant n. 226455
\emph{Supersymmetry, Quantum Gravity and Gauge Fields (Superfelds)}.
The work of A.S. was supported in part by the RFBR Grants No. 11-02-01335-a, No. 13-02-91330-NNIO-a and No. 13-02-90602-Arm-a.
%\newpage


\begin{thebibliography}{99}
%
\bibitem{Ade:2013uln}
P.~A.~R.~Ade {\it et al.}  [Planck Collaboration],
  \emph{Planck 2013 results. XXII. Constraints on inflation,}
  arXiv:1303.5082 [astro-ph.CO].
%
\bibitem{Ade:2013zuv}
  P.~A.~R.~Ade {\it et al.}  [Planck Collaboration],
  \emph{Planck 2013 results. XVI. Cosmological parameters,}
  arXiv:1303.5076 [astro-ph.CO].
%
  \bibitem{Hinshaw:2012aka}
  G.~Hinshaw {\it et al.}  [WMAP Collaboration],
  \emph{Nine-Year Wilkinson Microwave Anisotropy Probe (WMAP) Observations: Cosmological Parameter Results,}
  arXiv:1212.5226 [astro-ph.CO].
%
\bibitem{Encyclopaedia} J. Martin, C. Ringeval, V. Vennin,
\emph{Encyclopaedia Inflationaris}
arXiv:1303.3787v3 [astro-ph.CO]
%
\bibitem{minimalsergioKLP} S.~Ferrara, R.~Kallosh, A.~Linde and M.~Porrati,
  \emph{Minimal Supergravity Models of Inflation,} arXiv:1307.7696 [hep-th].
  To appear in Phys. Rev. D.
%
%
\bibitem{Ferrara:2013wka}
  S.~Ferrara, R.~Kallosh and A.~Van Proeyen,
  \emph{On the Supersymmetric Completion of $R+R^2$ Gravity and Cosmology},
  arXiv:1309.4052 [hep-th]. To appear in JHEP.
%
\bibitem{Ferrara:2013kca}
  S.~Ferrara, R.~Kallosh, A.~Linde and M.~Porrati,
  \emph{Higher Order Corrections in Minimal Supergravity Models of Inflation},
  arXiv:1309.1085 [hep-th]. To appear in JCAP.
  %
  \bibitem{VanProeyen:1979ks}
  A.~Van Proeyen,
  \emph{Massive Vector Multiplets in Supergravity,} Nucl.\ Phys.\ B {\bf 162} (1980) 376.
%%%%%%%%%%%%%%%%%%
%
\bibitem{Cecotti:1987qe}
  S.~Cecotti, S.~Ferrara, M.~Porrati and S.~Sabharwal,
  \emph{New Minimal Higher Derivative Supergravity Coupled To Matter,}
  Nucl.\ Phys.\ B {\bf 306} (1988) 160.
%%%%%%%%%%%%%%%%%
\bibitem{Dudasprimo}
 E.~Dudas, N.~Kitazawa and A.~Sagnotti,\emph{On climbing scalars in String Theory},
 Phys. Lett. B694 (2010) 80-88,
 arXiv:1009.0874 [hep-th].
%
 \bibitem{Dudas:2012vv}
  E.~Dudas, N.~Kitazawa, S.~P.~Patil and A.~Sagnotti,
  \emph{CMB Imprints of a Pre-Inflationary Climbing Phase,}
  JCAP {\bf 1205} (2012) 012,
  arXiv:1202.6630 [hep-th].
%
\bibitem{Sagnotti:2013ica}
  A.~Sagnotti,
  \emph{Brane SUSY Breaking and Inflation: Implications for Scalar Fields and CMB Distorsion,}
  arXiv:1303.6685 [hep-th].
 %
\bibitem{noicosmoitegr}
P.~Fr\'e, A.~Sagnotti and A.~S.~Sorin,
\emph{Integrable Scalar Cosmologies I. Foundations and links with String Theory},
  arXiv:1307.1910 [hep-th]. To appear in Nucl. Phys. B.
 %
\bibitem{mariosashapietrocosmo} P.~Fre, A.~S.~Sorin and M.~Trigiante,
  \emph{Integrable Scalar Cosmologies II. Can they fit into Gauged Extended Supergavity or be encoded in N=1 superpotentials?,}
  arXiv:1310.5340 [hep-th].
%
\bibitem{primosashapietro}
 P.~Fr\'e and A.~S.~Sorin,
  \emph{Inflation and Integrable one-field Cosmologies embedded in Rheonomic Supergravity,}
  arXiv:1308.2332 [hep-th]. To appear on Fortschritte der Physik.
%
\bibitem{piesashatwo} P.~Fre and A.~S.~Sorin,
  \emph{Axial Symmetric Kahler manifolds, the D-map of Inflaton Potentials and the Picard-Fuchs Equation,}
  arXiv:1310.5278 [hep-th]. To appear on Fortschritte der Physik.
%
%
\bibitem{Starobinsky:1980te}
  A.~A.~Starobinsky,
  \emph{A New Type of Isotropic Cosmological Models Without Singularity,}
  Phys.\ Lett.\ B {\bf 91} (1980) 99.
  %

%
\bibitem{johndimitri} J.~Ellis, D.~Nanopoulos, K.A.~Olive,
\emph{A No-Scale Supergravity Realization of the Starobinsky Model},
arXiv:1305.1247 [hep-th].
%
  \bibitem{Ketov:2010qz}
  S.~V.~Ketov and A.~A.~Starobinsky,
  \emph{Embedding $(R+R^2)$-Inflation into Supergravity,}
  Phys.\ Rev.\ D {\bf 83} (2011) 063512,
  arXiv:1011.0240 [hep-th].
%
\bibitem{Ketov:2012jt}
  S.~V.~Ketov and A.~A.~Starobinsky,
  \emph{Inflation and non-minimal scalar-curvature coupling in gravity and supergravity,}
  JCAP {\bf 1208} (2012) 022,
  arXiv:1203.0805 [hep-th].
%
%zdes'%%%%%%%%%%%%%%%%%%%%%%%%%%%%%%%%%%%%%%%%%
%
\bibitem{Kallosh:2013maa}
  R.~Kallosh and A.~Linde,
  \emph{Non-minimal Inflationary Attractors,}
  JCAP {\bf 1310} (2013) 033,
  arXiv:1307.7938 [hep-th].
%
\bibitem{Kallosh:2013daa}
  R.~Kallosh and A.~Linde,
  \emph{Multi-field Conformal Cosmological Attractors,}
  arXiv:1309.2015 [hep-th].
%
  \bibitem{Kallosh:2013tua}
  R.~Kallosh, A.~Linde and D.~Roest,
  \emph{A universal attractor for inflation at strong coupling,}
  arXiv:1310.3950 [hep-th].
%
 \bibitem{Kallosh:2013yoa}
  R.~Kallosh, A.~Linde and D.~Roest,
  \emph{Superconformal Inflationary $\alpha$-Attractors,}
  arXiv:1311.0472 [hep-th].
%
%%%%%%%%%%%%%%%%%%%%%%%%%%%%%%%%%%%%%%%%%%%%%%%%%%%%%
 %
  \bibitem{Kallosh:2013hoa}
  R.~Kallosh and A.~Linde,
  \emph{Universality Class in Conformal Inflation},
  JCAP {\bf 1307} (2013) 002,
  arXiv:1306.5220 [hep-th].
%
\bibitem{Kallosh:2013lkr}
  R.~Kallosh and A.~Linde,
  \emph{Superconformal generalizations of the Starobinsky model},
  JCAP {\bf 1306} (2013) 028,
  arXiv:1306.3214 [hep-th].
 %
 \bibitem{Farakos:2013cqa}
  F.~Farakos, A.~Kehagias and A.~Riotto,
  \emph{On the Starobinsky Model of Inflation from Supergravity},
  arXiv:1307.1137 [hep-th].
%

 %%%%%%%%%%%%%%%%%%%%%%%%%%%%%%%%%%%
\bibitem{Bagger:1982fn}
  J.~Bagger and E.~Witten,
  \emph{The Gauge Invariant Supersymmetric Nonlinear Sigma Model,} Phys.\ Lett.\ B {\bf 118} (1982) 103.
%
\bibitem{Fayet:1974jb}
  P.~Fayet and J.~Iliopoulos,
  \emph{Spontaneously Broken Supergauge Symmetries and Goldstone Spinors,}  Phys.\ Lett.\ B {\bf 51} (1974) 461.
%%%%%%%%%%%%%%%%%%%%%%%%%%%%%%%%%%%%
\bibitem{Komargodski:2009pc}
  Z.~Komargodski and N.~Seiberg,
  \emph{Comments on the Fayet-Iliopoulos Term in Field Theory and Supergravity,}
  JHEP {\bf 0906} (2009) 007
  [arXiv:0904.1159 [hep-th]].
%%%%%%%%%%%%%%%%%%%%%%%%%%%%%%%%%%%%%%
\bibitem{Freedman:1976uk}
  D.~Z.~Freedman,
  \emph{Supergravity with Axial Gauge Invariance,}  Phys.\ Rev.\ D {\bf 15} (1977) 1173.
%%%%%%%%%%%%%%%%%
\bibitem{D'Auria:1990fj}
  R.~D'Auria, S.~Ferrara and P.~Fre,
  \emph{Special and quaternionic isometries: General couplings in N=2 supergravity and the scalar potential,}  Nucl.\ Phys.\ B {\bf 359} (1991) 705.
%
\bibitem{standardN1} E. Cremmer, S. Ferrara, L. Girardello, A. Van Proeyen, \emph{Yang-Mills Theories with Local supersymmetry}, Nucl. Phys. \textbf{B212} (1983) 413-442.
%
\bibitem{castadauriafre2}
  L.~Castellani, R.~D'Auria and P.~Fre,
  \emph{Supergravity and superstrings: A Geometric perspective. Vol. 2: Supergravity,}
  Singapore, Singapore: World Scientific (1991) 607-1371.
%
\bibitem{NoistandardN2} L.~Andrianopoli, M.~Bertolini, A.~Ceresole, R.~D'Auria, S.~Ferrara, P.~Fre and T.~Magri,
  \emph{N=2 supergravity and N=2 superYang-Mills theory on general scalar manifolds: Symplectic covariance, gaugings and the momentum map}
  J.\ Geom.\ Phys.\  {\bf 23} (1997) 111
  [hep-th/9605032].
 %
 \bibitem{sasakimukhanov}
 V. Mukhanov, \emph{Physical foundations of cosmology,} Cambridge, UK: Univ. Pr. (2005) 421
 S. Weinberg, \emph{Cosmology},  Oxford, UK: Oxford Univ. Pr. (2008) 593 p;
 D. H. Lyth and A. R. Liddle, \emph{The primordial density perturbation: Cosmology, infation and the
origin of structure}, Cambridge, UK: Cambridge Univ. Pr. (2009) 497 p.
%
\bibitem{Cecotti:1987sa}
  S.~Cecotti,
  \emph{Higher Derivative Supergravity Is Equivalent To Standard Supergravity Coupled To Matter. 1},
  Phys.\ Lett.\ B {\bf 190}, 86 (1987).
%
%
\bibitem{contownsend} R. D'Auria,  P. Fr\'e, P. van Nieuwenhuizen, P. Townsend,
\emph{Invariance Of Actions, Rheonomy And The New Minimal N=1 Supergravity In The Group Manifold Approach}, Ann. Phys. \textbf{155} (1984) 423.
%
\bibitem{D'Auria:1988qm}
  R.~D'Auria, P.~Fre, G.~de Matteis and I.~Pesando,
  \emph{Superspace Constraints And Chern-simons Cohomology In D = 4 Superstring Effective Theories,}
  Int.\ J.\ Mod.\ Phys.\ A {\bf 4} (1989) 3577.
%
\bibitem{pietrosashasergioTwo} S. Ferrara, P. Fr\'e, A.S. Sorin, to appear.

%

\end{thebibliography}
\end{document}